\documentclass[prc,twocolumn,superscriptaddress,showpacs,amssymb,amsmath,amsfonts,aps]{revtex4}
\usepackage{graphicx}
\usepackage{dcolumn}
\usepackage{epsfig}

\begin{document}
\title{\large Electroexcitation of nucleon resonances
from CLAS data on single pion electroproduction\\ }

\newcommand*{\JLAB}{Thomas Jefferson National Accelerator Facility, Newport News, Virginia 23606}
\newcommand*{\JLABindex}{34}
\affiliation{\JLAB}
\newcommand*{\YEREVAN}{Yerevan Physics Institute, 375036 Yerevan, Armenia}
\newcommand*{\YEREVANindex}{39}
\affiliation{\YEREVAN}
\newcommand*{\FU}{Fairfield University, Fairfield CT 06824}
\newcommand*{\FUindex}{12}
\affiliation{\FU}
\newcommand*{\UNH}{University of New Hampshire, Durham, New Hampshire 03824-3568}
\newcommand*{\UNHindex}{26}
\affiliation{\UNH}
\newcommand*{\UCONN}{University of Connecticut, Storrs, Connecticut 06269}
\newcommand*{\UCONNindex}{10}
\affiliation{\UCONN}
\newcommand*{\VIRGINIA}{University of Virginia, Charlottesville, Virginia 22901}
\newcommand*{\VIRGINIAindex}{37}
\affiliation{\VIRGINIA}
\newcommand*{\KYUNGPOOK}{Kyungpook National University, Daegu 702-701, Republic of Korea}
\newcommand*{\KYUNGPOOKindex}{24}
\affiliation{\KYUNGPOOK}
\newcommand*{\RPI}{Rensselaer Polytechnic Institute, Troy, New York 12180-3590}
\newcommand*{\RPIindex}{30}
\affiliation{\RPI}
\newcommand*{\ODU}{Old Dominion University, Norfolk, Virginia 23529}
\newcommand*{\ODUindex}{29}
\affiliation{\ODU}
\newcommand*{\INFNGE}{INFN, Sezione di Genova, 16146 Genova, Italy}
\newcommand*{\INFNGEindex}{19}
\affiliation{\INFNGE}
\newcommand*{\SACLAY}{CEA, Centre de Saclay, Irfu/Service de Physique Nucl\'eaire, 
91191 Gif-sur-Yvette, France}
\newcommand*{\SACLAYindex}{8}
\affiliation{\SACLAY}
\newcommand*{\ITEP}{Institute of Theoretical and Experimental Physics, Moscow, 117259, Russia}
\newcommand*{\ITEPindex}{22}
\affiliation{\ITEP}
\newcommand*{\CMU}{Carnegie Mellon University, Pittsburgh, Pennsylvania 15213}
\newcommand*{\CMUindex}{6}
\affiliation{\CMU}
\newcommand*{\FSU}{Florida State University, Tallahassee, Florida 32306}
\newcommand*{\FSUindex}{14}
\affiliation{\FSU}
\newcommand*{\ANL}{Argonne National Laboratory, Argonne, Illinois 60441}
\newcommand*{\ANLindex}{1}
\affiliation{\ANL}
\newcommand*{\ASU}{Arizona State University, Tempe, Arizona 85287-1504}
\newcommand*{\ASUindex}{2}
\affiliation{\ASU}
\newcommand*{\UCLA}{University of California at Los Angeles, Los Angeles, California  90095-1547}
\newcommand*{\UCLAindex}{3}
\affiliation{\UCLA}
\newcommand*{\CSU}{California State University, Dominguez Hills, Carson, CA 90747}
\newcommand*{\CSUindex}{4}
\affiliation{\CSU}
\newcommand*{\Canisius}{Canisius College, Buffalo, NY}
\newcommand*{\Canisiusindex}{5}
\affiliation{\Canisius}
\newcommand*{\CUA}{Catholic University of America, Washington, D.C. 20064}
\newcommand*{\CUAindex}{7}
\affiliation{\CUA}
\newcommand*{\CNU}{Christopher Newport University, Newport News, Virginia 23606}
\newcommand*{\CNUindex}{9}
\affiliation{\CNU}
\newcommand*{\URICH}{University of Richmond, Richmond, Virginia 23173}
\newcommand*{\URICHindex}{40}
\affiliation{\URICH}
\newcommand*{\ECOSSEE}{Edinburgh University, Edinburgh EH9 3JZ, United Kingdom}
\newcommand*{\ECOSSEEindex}{11}
\affiliation{\ECOSSEE}
\newcommand*{\FIU}{Florida International University, Miami, Florida 33199}
\newcommand*{\FIUindex}{13}
\affiliation{\FIU}
\newcommand*{\GWU}{The George Washington University, Washington, DC 20052}
\newcommand*{\GWUindex}{15}
\affiliation{\GWU}
\newcommand*{\ECOSSEG}{University of Glasgow, Glasgow G12 8QQ, United Kingdom}
\newcommand*{\ECOSSEGindex}{16}
\affiliation{\ECOSSEG}
\newcommand*{\ISU}{Idaho State University, Pocatello, Idaho 83209}
\newcommand*{\ISUindex}{17}
\affiliation{\ISU}
\newcommand*{\INFNFR}{INFN, Laboratori Nazionali di Frascati, 00044 Frascati, Italy}
\newcommand*{\INFNFRindex}{18}
\affiliation{\INFNFR}
\newcommand*{\INFNRO}{INFN, Sezione di Roma Tor Vergata, 00133 Rome, Italy}
\newcommand*{\INFNROindex}{20}
\affiliation{\INFNRO}
\newcommand*{\ORSAY}{Institut de Physique Nucl\'eaire ORSAY, Orsay, France}
\newcommand*{\ORSAYindex}{21}
\affiliation{\ORSAY}
\newcommand*{\JMU}{James Madison University, Harrisonburg, Virginia 22807}
\newcommand*{\JMUindex}{23}
\affiliation{\JMU}
\newcommand*{\MIT}{Massachusetts Institute of Technology, Cambridge, Massachusetts  02139-4307}
\newcommand*{\MITindex}{25}
\affiliation{\MIT}
\newcommand*{\NSU}{Norfolk State University, Norfolk, Virginia 23504}
\newcommand*{\NSUindex}{27}
\affiliation{\NSU}
\newcommand*{\OHIOU}{Ohio University, Athens, Ohio  45701}
\newcommand*{\OHIOUindex}{28}
\affiliation{\OHIOU}
\newcommand*{\ROMAII}{Universita' di Roma Tor Vergata, 00133 Rome Italy}
\newcommand*{\ROMAIIindex}{31}
\affiliation{\ROMAII}
\newcommand*{\MOSCOW}{Skobeltsyn Nuclear Physics Institute, 119899 
Moscow, Russia}
\newcommand*{\MOSCOWindex}{32}
\affiliation{\MOSCOW}
\newcommand*{\SCAROLINA}{University of South Carolina, Columbia, South Carolina 29208}
\newcommand*{\SCAROLINAindex}{33}
\affiliation{\SCAROLINA}
\newcommand*{\UNIONC}{Union College, Schenectady, NY 12308}
\newcommand*{\UNIONCindex}{35}
\affiliation{\UNIONC}
\newcommand*{\UTFSM}{Universidad T\'{e}cnica Federico Santa Mar\'{i}a, Casilla 110-V Valpara\'{i}so, 
Chile}
\newcommand*{\UTFSMindex}{36}
\affiliation{\UTFSM}
\newcommand*{\WM}{College of William and Mary, Williamsburg, Virginia 23187-8795}
\newcommand*{\WMindex}{38}
\affiliation{\WM}
\newcommand*{\NOWLPSC}{LPSC-Grenoble, France}
\newcommand*{\NOWLANL}{Los Alamos National Laborotory, New Mexico, NM}
\newcommand*{\NOWGWU}{The George Washington University, Washington, DC 20052}
\newcommand*{\NOWJLAB}{Thomas Jefferson National Accelerator Facility, Newport News, Virginia 23606}
\newcommand*{\NOWCNU}{Christopher Newport University, Newport News, Virginia 23606}
\newcommand*{\NOWECOSSEE}{Edinburgh University, Edinburgh EH9 3JZ, United Kingdom}
\newcommand*{\NOWWM}{College of William and Mary, Williamsburg, Virginia 23187-8795}
\author{I.G.~Aznauryan}
     \affiliation{\JLAB}
     \affiliation{\YEREVAN}
\author{V.D.~Burkert}
     \affiliation{\JLAB}
\author {A.S.~Biselli} 
\affiliation{\FU}
\author {H.~Egiyan} 
\affiliation{\JLAB}
\affiliation{\UNH}
\author{K.~Joo}
     \affiliation{\UCONN}
     \affiliation{\VIRGINIA}
\author{W.~Kim}
     \affiliation{\KYUNGPOOK}
\author {K.~Park} 
    \affiliation{\JLAB}
\affiliation{\KYUNGPOOK}
\author{L.C.~Smith}
     \affiliation{\VIRGINIA}
\author {M.~Ungaro} 
\affiliation{\JLAB}
\affiliation{\UCONN}
\affiliation{\RPI}
\author {K. P. ~Adhikari} 
\affiliation{\ODU}
% ...
 
\collaboration{The CLAS Collaboration}
     \noaffiliation

\begin{abstract}
{We present results
on the electroexcitation
of the low mass resonances 
$\Delta(1232)P_{33}$, $N(1440)P_{11}$, 
$N(1520)D_{13}$, and
$N(1535)S_{11}$ in a wide range of  $Q^2$. 
The results were obtained in the comprehensive analysis
of JLab-CLAS  data on differential cross sections, longitudinally 
polarized
beam asymmetries, and longitudinal target and beam-target asymmetries 
for 
$\pi$ electroproduction off the proton.
The data were analysed  using two conceptually
different approaches, fixed-$t$ dispersion relations
and a unitary isobar model, 
allowing us to draw conclusions on the  model sensitivity
of the obtained electrocoupling amplitudes.
The amplitudes for the $\Delta(1232)P_{33}$ show the importance of
a meson-cloud contribution to quantitatively explain the magnetic
dipole
strength, as well as the electric and scalar quadrupole transitions.
They do not show any tendency of approaching the pQCD regime for
$Q^2\leq 6~$GeV$^2$.  For
the Roper resonance, $N(1440)P_{11}$, the data provide strong evidence
for this
state as a predominantly radial excitation of a 3-quark
ground state. 
Measured in pion electroproduction,
the transverse helicity amplitude
for the $N(1535)S_{11}$ allowed us 
to obtain the branching ratios of this state
to the $\pi N$ and $\eta N$ channels
via comparison
to the results extracted from $\eta$ electroproduction.
The extensive CLAS data also enabled the extraction of
the $\gamma^*p\rightarrow
N(1520)D_{13}$ and $N(1535)S_{11}$ longitudinal
helicity amplitudes with good precision. 
For the $N(1535)S_{11}$,  
these  results 
became a challenge for quark models,
and may be indicative of
large meson-cloud contributions or
of representations of this state different from a 3q excitation.
The transverse amplitudes for the $N(1520)D_{13}$ clearly show the 
rapid changeover from
helicity-3/2 dominance at the real photon point to
helicity-1/2
dominance at $Q^2 > 1~$GeV$^2$, confirming a long-standing prediction
of the constituent quark model.
}
\end{abstract}
\maketitle
\section{Introduction}

The excitation of nucleon resonances in electromagnetic
interactions has long been recognized as an important source
of information to understand the 
strong interaction in the domain of quark confinement.
The CLAS detector at 
Jefferson Lab
is the first large acceptance instrument
designed for the comprehensive investigation
of exclusive electroproduction of mesons with the goal
to study the electroexcitation of nucleon
resonances in detail. 
In recent years, a variety of measurements
of single pion electroproduction on protons,
including polarization measurements, have been performed
at CLAS in a wide range of photon virtuality $Q^2$ from 0.16 to 6 GeV$^2$
\cite{Joo1,Joo2,Joo3,Egiyan,Ungaro,Smith,Park,Biselli}.
In this work we present the results on the 
electroexcitation of the resonances  
$\Delta(1232)P_{33}$, $N(1440)P_{11}$,
$N(1520)D_{13}$, and
$N(1535)S_{11}$, obtained from the
comprehensive analysis of these data.

Theoretical and experimental investigations
of the electroexcitation of nucleon resonances
have a long history, and 
along with the hadron masses and nucleon electromagnetic
characteristics, the information on the
$\gamma^*N\rightarrow N^*$ transitions played
an important role in the justification of the quark model.
However, the picture of the nucleon and its excited states,
which at first seemed quite simple and was
identified as a model of non-relativistic constituent quarks,
turned out to be more complex.
One of the reasons for this was the realization
that quarks are relativistic objects.
A consistent way to perform the relativistic treatment
of the $\gamma^*N\rightarrow N(N^*)$ transitions
is to consider them in the light-front (LF) dynamics
\cite{Drell,Terentev,Brodsky}.
The relevant approaches were developed and used
to describe the nucleon and its excited states 
\cite{Aznquark,Aznquark1,Weber,Capstick,Simula,Simula1,Bruno,AznRoper}. 
However, much more effort is required to obtain a  
better understanding of what are the
$N$ and $N^*$ LF wave functions
and what is their connection to
the inter-quark forces and 
to the QCD confining mechanism.
Another reason is connected with the realization
that the traditional picture of baryons built
from three constituent quarks is an oversimplified
approximation.
In the case of the
$N(1440)P_{11}$ and $N(1535)S_{11}$, the mass ordering
of these states, the large total width of $N(1440)P_{11}$,
and the substantial
coupling of $N(1535)S_{11}$ to the $\eta N$ channel \cite{PDG}
and to strange particles \cite{Liu,Xie},
are indicative of posible additional
$q\bar{q}$ components in the wave functions
of these states \cite{Riska,An} and (or) of alternative
descriptions.
Within dynamical reaction models \cite{Yang,Kamalov,Sato,Lee}, the 
meson-cloud
contribution is identified as a
source of the long-standing discrepancy
between the data and constituent quark model 
predictions for the $\gamma^*N\rightarrow \Delta(1232)P_{33}$
magnetic-dipole amplitude.
The importance of pion (cloud) contributions
to the transition form factors
has also been confirmed by the lattice calculations \cite{Alexandrou}.
Alternative descriptions include
the representation
of $N(1440)P_{11}$ as a gluonic baryon excitation 
\cite{Li1,Li2} and the possibility
that nucleon resonances
are meson-baryon molecules  
generated in chiral coupled-channel
dynamics \cite{Weise,Krehl,Nieves,Oset1,Lutz}. 
Relations between baryon electromagnetic form factors and 
generalized
parton distributions (GPDs) have also been formulated that
connect these two different notions to describe the
baryon structure \cite{GPD1,GPD2}.

The improvement
in accuracy and reliability of the information
on the electroexcitation of the nucleon's excited states  
over a large range in photon virtuality $Q^2$
is very important for the progress in our understanding of 
this complex picture of the strong interaction in the domain
of quark confinement. 

Our goal is to determine in detail the $Q^2$-behavior
of the electroexcitation of resonances. For this reason,
we analyse the data at each $Q^2$ point separately without 
imposing any constraints on the $Q^2$ dependence of the 
electroexcitation amplitudes. This is in contrast
with the analyses by MAID, for instance
MAID2007 \cite{MAID}, where the  
electroexcitation amplitudes are in part constrained 
by using parameterizations for their $Q^2$ dependence. 

The analysis  was performed using
two approaches,  fixed-$t$ dispersion relations (DR) and 
the unitary isobar model (UIM). The real parts of
the amplitudes, which contain a significant part
of the non-resonant contributions, are built in these approaches  
in conceptually different ways. 
This allows us to draw conclusions on the model
sensitivity of the  resulting electroexcitation amplitudes.

The paper is organized as follows. 
In Sec. II, we present the data and 
discuss the 
stages of the analysis.
The approaches we use to analyse the data,
DR and UIM,  
were successfully employed 
in analyses of pion-photoproduction
and low-$Q^2$-electroproduction data, 
see Refs. \cite{Azn0,Azn04,Azn065}. 
In  Sec. III we therefore discuss only the points that 
need different treatment when we move from 
low $Q^2$ to high $Q^2$. 
Uncertainties 
of the background contributions
related to
the pion and nucleon elastic form factors,  and to 
$\rho,\omega\rightarrow \pi\gamma$ transition form factors
are discussed in Sec. IV.
In Sec. V, we present how resonance contributions
are taken into account and explain
how the uncertainties associated with 
higher resonances 
and with the uncertainties of masses and widths
of the  $N(1440)P_{11}$,
$N(1520)D_{13}$, and $N(1535)S_{11}$ 
are accounted for.
All these uncertainties are included
in the total model uncertainty of the final results.
So, in addition to the uncertainties in the data,
we have accounted for, as much as possible,
the model uncertainties of the 
extracted  $\gamma^*N\rightarrow~\Delta(1232)P_{33}$, 
$N(1440)P_{11}$, 
$N(1520)D_{13}$, and 
$N(1535)S_{11}$ amplitudes.  
The results are presented in Sec. VI, compared with
model predictions in Sec. VII, and summarized in 
Sec. VIII.

\section{Data analysis considerations}

The data are presented in Tables 
\ref{pol_data}-\ref{pi0_data}.
They
cover the first, second,
and part of the third resonance regions. 
The stages of our analysis are dictated
by how we evaluate the influence
of higher resonances 
on the extracted amplitudes
for the $\Delta(1232)P_{33}$ 
and for the resonances 
from the second resonance region.

In the first stage, we analyse the
data reported in Table \ref{pol_data}   
($Q^2=0.3-0.65~$GeV$^2$)
where the richest set of polarization measurements
is available.
The results based  
on the analysis of the cross sections and  longitudinally polarized
beam asymmetries ($A_{LT'}$) 
at $Q^2=0.4$ and $0.6-0.65~$GeV$^2$
were already presented in Refs. \cite{Azn04,Azn065}.
However, recently, new data 
have become available 
from the JLab-CLAS measurements of
longitudinal target ($A_t$) and beam-target ($A_{et}$) asymmetries for
$\vec{e}\vec{p}\rightarrow ep\pi^0$ at $Q^2=0.252,~0.385,~0.611~$
GeV$^2$ 
\cite{Biselli}. For this reason,
we performed a new analysis on the same
data set,  including these 
new measurements. We also extended our analysis 
to the available
data for the close values of $Q^2=0.3$ and $0.5-0.525~$GeV$^2$.
As the asymmetries $A_{LT'},~A_t,~A_{et}$
have relatively weak $Q^2$ dependences, the data on asymmetries
at nearby $Q^2$ were also included in the corresponding sets 
at $Q^2=0.3$ and $0.5-0.525~$GeV$^2$. 
Following our previous analyses \cite{Azn04,Azn065},
we have complemented the data set 
at $Q^2=0.6-0.65~$GeV$^2$ with the DESY  
$\pi^+$ cross sections data \cite{Alder},
since the corresponding CLAS data extend over a restricted
range in $W$.  

In Ref. \cite{Azn065}, the analysis of data 
at $Q^2=0.6-0.65~$GeV$^2$ was
performed in combination
with JLab-CLAS data for double-pion
electroproduction off the proton \cite{Fedotov}. This allowed us to get 
information
on the electroexcitation amplitudes for the resonances
from the third resonance region.
This information, combined with the $\gamma p \rightarrow N^*$
amplitudes known from photoproduction data \cite {PDG},
sets the ranges of the higher resonance contributions
when we extract the amplitudes of the $\gamma^* p \rightarrow$ 
$\Delta(1232)P_{33}$,
$N(1440)P_{11}$,
$N(1520)D_{13}$, and $N(1535)S_{11}$ transitions 
from the data reported in Table \ref{pol_data}.

In the next step, we analyse the data from Table 
\ref{pip_data}
which present a large body of
$\vec{e}p\rightarrow en\pi^+$
differential
cross sections and longitudinally polarized electron beam
asymmetries
at large $Q^2=1.72-4.16~$GeV$^2$
\cite{Park}. As the isospin $\frac{1}{2}$ nucleon
resonances couple more strongly
to the $\pi^+ n$ channel, these data provide large sensitivity 
to the electrocouplings of the 
$N(1440)P_{11}$, $N(1520)D_{13}$, and $N(1535)S_{11}$
states. 
Until recently, the information on the electroexcitation
of these resonances
at $Q^2>1~$GeV$^2$ was based almost exclusively
on the (unpublished) DESY data \cite{Haidan}
on $ep\rightarrow ep\pi^0$
($Q^2\approx 2$ and $3~$GeV$^2$) which have very limited
angular coverage. Furthermore, the $\pi^0 p$ final state 
is coupled more weakly to the  isospin $\frac{1}{2}$ states,
and is dominated by the nearby
isospin $\frac{3}{2}$ 
$\Delta(1232)P_{33}$
resonance.
For the $N(1535)S_{11}$, which has a large branching
ratio to the $\eta N$ channel, there is also information
on the
$\gamma^*N\rightarrow N(1535)S_{11}$
transverse helicity amplitude
found from the data on $\eta$ electroproduction
off the proton 
\cite{Armstrong,Thompson,Denizli}.

In the range of $Q^2$ covered by the data \cite{Park}
(Table \ref{pip_data}),
there is no information on the helicity amplitudes
for the resonances 
from the third resonance region.
The data \cite{Park} 
cover only part of this region
and do not allow us
to extract reliably the corresponding amplitudes
(except those for $N(1680)F_{15}$).
For the 
$\gamma^* p\rightarrow N(1440)P_{11}$, 
$N(1520)D_{13}$, and $N(1535)S_{11}$
amplitudes extracted from the data \cite{Park}, 
the  evaluation of
the uncertainties 
caused by the lack of information on the resonances 
from the third resonance region
is described in Sec. V.

Finally, we extract the 
$\gamma^* p\rightarrow \Delta(1232)P_{33}$ 
amplitudes  from the data reported
in Tables \ref{smith} and \ref{pi0_data}.
These are low $Q^2$ data for $\pi^0$ and $\pi^+$ 
electroproduction differential
cross sections \cite{Smith}  and data for
$\pi^0$ electroproduction differential
cross sections  at 
$Q^2=1.15,1.45~$GeV$^2$ \cite{Joo1} 
and $3-6~$GeV$^2$ \cite{Ungaro}. 
In the analysis of these data, the influence
of higher resonances on the results for the $\Delta(1232)P_{33}$
was evaluated by 
employing the spread of the $\gamma^* p \rightarrow N(1440)P_{11}$,
$N(1520)D_{13}$, and $N(1535)S_{11}$ amplitudes obtained
in the previous stages of our analysis of the data 
from Tables \ref{pol_data} and \ref{pip_data}.

Although the data for $Q^2=0.75-1.45~$GeV$^2$
(Table \ref{pi0_data})
cover a wide range in $W$, the
absence of $\pi^+$ electroproduction data for 
these $Q^2$, except $Q^2=0.9~$GeV$^2$, does not allow us to 
extract the amplitudes for the 
$N(1440)P_{11}$, $N(1520)D_{13}$, $N(1535)S_{11}$
resonances with model uncertainties comparable to those
for the amplitudes found from the data  
of Tables \ref{pol_data} and \ref{pip_data}.
For $Q^2\simeq 0.95~$GeV$^2$, there are DESY
$\pi^+$ electroproduction data \cite{Alder},
which cover the second and third resonance regions,
allowing us to extract amplitudes for 
all resonances
from the first and second resonance regions
at $Q^2=0.9-0.95~$GeV$^2$. 
To evaluate
the uncertainties 
caused by the higher mass resonances, 
we have used 
for $Q^2=0.9-0.95~$GeV$^2$ 
the same procedure as for the data  
from Table \ref{pip_data}.

\begin{table}[t]
\begin{center}
\begin{tabular}{|cccccccc|}
\hline
&&&Number&&&&\\
&&&of data&&$\frac{\chi^2}{N}$&&\\
Obser-&$Q^2$&$W$&points&&&&Ref.\\
vable&(GeV$^2$)&(GeV)&($N$)&DR&${}$&UIM&\\
\hline
$\frac{d\sigma}{d\Omega}(\pi^+)$&0.3&1.1-1.55&2364&2.06&&1.93
&\cite{Egiyan}\\
$A_{t}(\pi^0)$&0.252&1.125-1.55&594&1.36&&1.48&\cite{Biselli}\\
$A_{et}(\pi^0)$&0.252&1.125-1.55&598&1.19&&1.23&\cite{Biselli}\\
\hline
$\frac{d\sigma}{d\Omega}(\pi^0)$&0.4&1.1-1.68
&3530&1.23&&1.24&\cite{Joo1}\\
$\frac{d\sigma}{d\Omega}(\pi^+)$&0.4&1.1-1.55&2308&1.92&&1.64
&\cite{Egiyan}\\
$A_{LT'}(\pi^0)$&0.4&1.1-1.66&956&1.24&&1.18&\cite{Joo2}\\
$A_{LT'}(\pi^+)$&0.4&1.1-1.66&918&1.28&&1.19&\cite{Joo3}\\
$A_{t}(\pi^0)$&0.385&1.125-1.55&696&1.40&&1.61&\cite{Biselli}\\
$A_{et}(\pi^0)$&0.385&1.125-1.55&692&1.22&&1.25&\cite{Biselli}\\
\hline
$\frac{d\sigma}{d\Omega}(\pi^0)$&0.525&1.1-1.66
&3377&1.33&&1.35&\cite{Joo1}\\
$\frac{d\sigma}{d\Omega}(\pi^+)$&0.5&1.1-1.51&2158&1.51&&1.48
&\cite{Egiyan}\\
\hline
$\frac{d\sigma}{d\Omega}(\pi^0)$&0.65&1.1-1.68
&6149&1.09&&1.14&\cite{Joo1}\\
$\frac{d\sigma}{d\Omega}(\pi^+)$&0.6&1.1-1.41&1484&1.21&&1.24
&\cite{Egiyan}\\
$\frac{d\sigma}{d\Omega}(\pi^+)$&$\simeq 0.6$ 
&1.4-1.76&477&1.72&&1.74&\cite{Alder}\\
$A_{LT'}(\pi^0)$&0.65&1.1-1.66&805&1.09&&1.13&\cite{Joo2}\\
$A_{LT'}(\pi^+)$&0.65&1.1-1.66&812&1.09&&1.04&\cite{Joo3}\\
$A_{t}(\pi^0)$&0.611&1.125-1.55&930&1.38&
&1.40&\cite{Biselli}\\
$A_{et}(\pi^0)$&0.611&1.125-1.55&923&1.26&
&1.28&\cite{Biselli}\\
\hline
\end{tabular}
\caption{\label{pol_data} The data sets included in the
first stage of the analysis, as discussed in the text.
The columns corresponding to DR and UIM show the
results for $\chi^2$ per data point obtained, respectively, using
fixed-$t$ dispersions relations and the unitary isobar model
described in Sec. III. 
}
\end{center}
\end{table}

\begin{table}[t]
\begin{center}
\begin{tabular}{|ccccccc|}
\hline
&&&Number of&&$\chi^2/N$&\\
Obser-&$Q^2$&$W$&data points&&&\\
vable&(GeV$^2$)&(GeV)&($N$)&DR&${}$&UIM\\
\hline
$\frac{d\sigma}{d\Omega}(\pi^+)$&1.72&1.11-1.69&3530&2.3&&2.5\\
&2.05&1.11-1.69&5123&2.3&&2.2\\
&2.44&1.11-1.69&5452&2.0&&2.0\\
&2.91&1.11-1.69&5484&1.9&&2.1\\
&3.48&1.11-1.69&5482&1.3&&1.4\\
&4.16&1.11-1.69&5778&1.1&&1.1\\
$A_{LT'}(\pi^+)$&1.72&1.12-1.68&699&2.9&&3.0\\
&2.05&1.12-1.68&721&3.0&&2.9\\
&2.44&1.12-1.68&725&3.0&&3.0\\
&2.91&1.12-1.68&767&2.7&&2.7\\
&3.48&1.12-1.68&623&2.4&&2.3\\
\hline
\end{tabular}
\caption{\label{pip_data} 
The $\vec{e}p\rightarrow en\pi^+$ data from Ref. \cite{Park}.}
\end{center}
\end{table}

\begin{table}[t]
\begin{center}
\begin{tabular}{|ccccccc|}
\hline
&&&Number of&&$\chi^2/N$&\\
Obser-&$Q^2$&$W$&data points&&&\\
vable&(GeV$^2$)&(GeV)&($N$)&DR&${}$&UIM\\
\hline
$\frac{d\sigma}{d\Omega}(\pi^0)$&0.16&1.1-1.38&3301&1.96&&1.98\\
$\frac{d\sigma}{d\Omega}(\pi^+)$&0.16&1.1-1.38&2909&1.69&&1.67\\
$\frac{d\sigma}{d\Omega}(\pi^0)$&0.20&1.1-1.38&3292&2.29&&2.24\\
$\frac{d\sigma}{d\Omega}(\pi^+)$&0.20&1.1-1.38&2939&1.76&&1.78\\
$\frac{d\sigma}{d\Omega}(\pi^0)$&0.24&1.1-1.38&3086&1.86&&1.82\\
$\frac{d\sigma}{d\Omega}(\pi^+)$&0.24&1.1-1.38&2951&1.49&&1.46\\
$\frac{d\sigma}{d\Omega}(\pi^0)$&0.28&1.1-1.38&2876&1.56&&1.59\\
$\frac{d\sigma}{d\Omega}(\pi^+)$&0.28&1.1-1.38&2941&1.47&&1.44\\
$\frac{d\sigma}{d\Omega}(\pi^0)$&0.32&1.1-1.38&2836&1.51&&1.48\\
$\frac{d\sigma}{d\Omega}(\pi^+)$&0.32&1.1-1.38&2922&1.39&&1.37\\
$\frac{d\sigma}{d\Omega}(\pi^0)$&0.36&1.1-1.38&2576&1.46&&1.42\\
$\frac{d\sigma}{d\Omega}(\pi^+)$&0.36&1.1-1.38&2611&1.35&&1.38\\
\hline
\end{tabular}
\caption{\label{smith} The low $Q^2$ 
data from Ref. \cite{Smith} analysed in the third stage
of the analysis. 
}
\end{center}
\end{table}

\begin{table}[t]
\begin{center}
\begin{tabular}{|ccccccc|}
\hline
&&&Number of&&$\chi^2/N$&\\
Obser-&$Q^2$&$W$&data points&&&\\
vable&(GeV$^2$)&(GeV)&($N$)&DR&${}$&UIM\\
\hline
$\frac{d\sigma}{d\Omega}(\pi^0)$&0.75&1.1-1.68&3555&1.16&&1.18\\
\hline
$\frac{d\sigma}{d\Omega}(\pi^0)$&0.9&1.1-1.68&3378&1.22&&1.25\\
$\frac{d\sigma}{d\Omega}(\pi^+)$&$\simeq 0.95$ 
&1.36-1.76&725&1.62&&1.66\\
\hline
$\frac{d\sigma}{d\Omega}(\pi^0)$&1.15&1.1-1.68&1796&1.09&&1.15\\
&1.45&1.1-1.62&1878&1.15&&1.18\\
&3&1.11-1.39&1800&1.41&&1.37\\
&3.5&1.11-1.39&1800&1.22&&1.24\\
&4.2&1.11-1.39&1800&1.16&&1.19\\
&5&1.11-1.39&1800&0.82&&0.88\\
&6&1.11-1.39&1800&0.66&&0.67\\
\hline
\end{tabular}
\caption{\label{pi0_data} The 
data included in the  third stage
of the analysis: the data for 
$\frac{d\sigma}{d\Omega}(\pi^0)$
at $Q^2=0.75-1.45$
and $3-6~$GeV$^2$
are from Refs. \cite{Joo1} and \cite{Park}, respectively;
the data for 
$\frac{d\sigma}{d\Omega}(\pi^+)$
are from Ref. \cite{Alder}.
}
\end{center}
\end{table}

\section{Analysis approaches}

The approaches we use to analyse the data,
DR and UIM, are 
described in detail in Refs. \cite{Azn0,Azn04}
and  were successfully employed in Refs. 
\cite{Azn0,Azn04,Azn065} for the analyses of pion-photoproduction
and low-$Q^2$-electroproduction data. 
In this Section we discuss certain aspects in these 
approaches that need a different treatment as we move  
to higher $Q^2$. 

\subsection{Dispersion relations}

We use fixed-$t$ dispersion relations for invariant 
amplitudes defined 
in accordance with the following definition of the
electromagnetic current
$I^{\mu}$ for the $\gamma^* N\rightarrow \pi N$ process \cite{Devenish}:
\begin{eqnarray}
&I^{\mu}\equiv\bar{u}(p_2)\gamma _5 
{\cal{I}}^{\mu}u(p_1)\phi_{\pi},\\ 
&{\cal{I}}^{\mu} =\frac{B_1}{2}\left[ \gamma
^\mu k\hspace{-1.8mm}\slash-
k\hspace{-1.8mm}\slash\gamma ^\mu \right]+2P ^\mu
B_2+2q^\mu B_3\\
&+2k^\mu B_4-\gamma ^\mu B_5+k\hspace{-1.8mm}\slash P^\mu B_6+
k\hspace{-1.8mm}\slash k^\mu B_7+
k\hspace{-1.8mm}\slash q^\mu B_8,\nonumber
\end{eqnarray}
where $k,~q,~p_1,~p_2$ are the four-momenta
of the virtual photon, pion, and initial and final nucleons, respectively;
$P=\frac{1}{2}(p_1+p_2),~B_1(s,t,Q^2),B_2(s,t,Q^2),...B_8(s,t,Q^2)$
are the invariant amplitudes that are functions
of the invariant variables
$s=(k+p_1)^2,~t=(k-q)^2,~Q^2\equiv
-k^2$; $u(p_1)$, $u(p_2)$ are the Dirac spinors of the initial
and final state nucleon, and $\phi_{\pi}$ is the pion field.
                                                                                                     
The conservation of  $I^{\mu}$
leads to the relations:
\begin{eqnarray}
&&4Q^2B_4=(s-u)B_2 -2(t+Q^2-m_{\pi} ^2)B_3, \\
&&2Q^2B_7=-2B'_5 -(t+Q^2-m_{\pi} ^2)B_8,
\label{2}\end{eqnarray}
where
$B'_5\equiv B_5-\frac{1}{4}(s-u)B_6$. Therefore,
only six of the eight invariant amplitudes
are independent.
In Ref. \cite{Azn0}, the following
independent amplitudes were chosen:
$B_1,B_2,B_3,B'_5,B_6,B_8$.
Taking into account the isotopic structure,
we have 18 independent invariant amplitudes.
For the amplitudes $B_1^{(\pm,0)},B_2^{(\pm,0)},B_3^{(+,0)},
{B'}_5^{(\pm,0)},B_6^{(\pm,0)},B_8^{(\pm,0)}$,
unsubtracted dispersion relations at fixed $t$ can be written.
The only exception is the amplitude $B_3^{(-)}$,
for which a subtraction is neccessary:
\begin{eqnarray}
&&Re~ 
B_3^{(-)}(s,t,Q^2)=f_{sub}(t,Q^2)
-ge\frac{F_{\pi}(Q^2)}{t-m_{\pi}^2}\nonumber\\
&&-\frac{ge}{4}\left[F_1^p(Q^2)-F_1^n(Q^2)\right]
\left(\frac{1}{s-m^2}+\frac{1}{u-m^2}\right)\\
&&+\frac{P}{\pi } \int \limits_{s_{thr}}^{\infty}
Im~ B_3^{(-)}(s',t,Q^2)
\left(\frac{1}{s'-s}+\frac{1 }{s'-u}\right)ds',\nonumber
\end{eqnarray}
where $g^2/4\pi=13.8$, $e^2/4\pi=1/137$, $F_{\pi}(Q^2)$
is the pion form factor, $F_1^N(Q^2)$ is the nucleon
Pauli form factor, and $m$ and $m_{\pi}$ are the
nucleon and pion masses, respectively.

At $Q^2=0$, using the relation $B_3=B_2\frac{s-u}{2(t-m_{\pi}^2)}$,
which follows from Eq. (3), and DR
for the amplitude $B_2(s,t,Q^2=0)$, 
one obtains:
\begin{equation}
f_{sub}(t,Q^2)=
4\frac{P}{\pi } \int \limits_{s_{thr}}^{\infty}
\frac{Im~ B_3^{(-)}(s',t,Q^2)}{u'-s'}ds',
\end{equation}
where $u'=2m^2+m_{\pi}^2-Q^2-s'-t$.

\begin{figure}[htp]
\begin{center}
\includegraphics[width=5.cm]{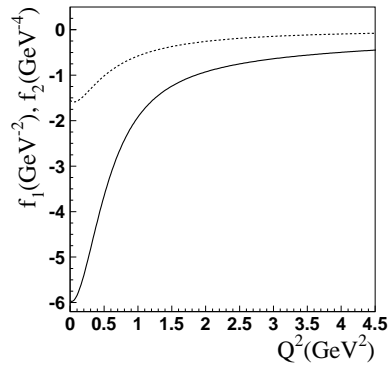}
\vspace{-0.1cm}
\caption{\small $Q^2$ dependence of the coefficients 
$f_1(Q^2)$ (solid curve) and $f_2(Q^2)$ (dashed curve) from Eq. (7).
The results for 
$Q^{2}< 0.7~$GeV$^{2}$  were found using Eq. (6), whereas
the results for 
$Q^{2}=1.7-4.5~$GeV$^{2}$  are from
the fit to the data \cite{Park}. \label{coef}}
\end{center}
\end{figure}

This expression for $f_{sub}(t,Q^2)$ was successfully used
for the analysis of pion photoproduction and 
low $Q^2=$0.4, 0.65 GeV$^2$
electroproduction data \cite{Azn0,Azn04}.
However, it turned out that it is not suitable 
at higher $Q^2$.
Using a simple parametrization:
\begin{equation}
f_{sub}(t,Q^2)=f_1(Q^2)+f_2(Q^2)t,
\end{equation}
a  suitable subtraction
was found from the fit to 
the data for $Q^2=1.7-4.5~$GeV$^2$ \cite{Park}.
The linear parametrization in $t$ is also consistent
with the subtraction found from Eq. (6) at low $Q^2$. 
Fig. \ref{coef} 
demonstrates smooth transition of the results for 
the coefficients $f_1(Q^2),f_2(Q^2)$
found at low 
$Q^{2}< 0.7~$GeV$^{2}$  using Eq. (6)
to those at large $Q^{2}=1.7-4.5~$GeV$^{2}$ found from
the fit to the data \cite{Park}.
                                                               
\begin{figure}[htp]
\begin{center}
\includegraphics[width=7.5cm]{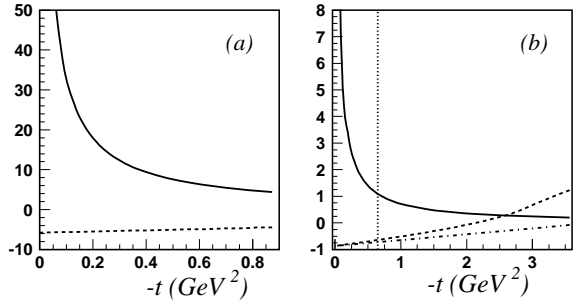}
\vspace{-0.1cm}
\caption{\small 
The pion contribution in GeV$^{-2}$ units (solid curves) 
to the DR for the
amplitude $B_3^{(-)}(s,t,Q^2)$, Eq. (5), compared to 
$f_{sub}(t,Q^2)$ at $Q^2=0$ (a) and $Q^2=2.44~$GeV$^2$ (b).
The dashed curves represent $f_{sub}(t,Q^2)$ taken in the form of Eq. (6),
the dash-dotted curve corresponds to the results for $f_{sub}(t,Q^2)$ 
obtained by fitting the data \cite{Park}. 
At $Q^2=2.44~$GeV$^2$, the physical
region is located on the right side of the 
dotted vertical line. 
\label{sub}}
\end{center}
\end{figure}

Fig. \ref{sub} shows the relative contribution
of $f_{sub}(t,Q^2)$ compared with the pion contribution
in Eq. (5)
at $Q^2=0$
and $Q^2=2.44~$GeV$^2$. It can be seen that the contribution
of $f_{sub}(t,Q^2)$ is comparable with the pion contribution
only at large $|t|$, where the latter is small.
At small $|t|$, $f_{sub}(t,Q^2)$ is very
small compared to the pion contribution.
                                                               
\subsection{Unitary isobar model}

The UIM 
of Ref. \cite{Azn0} was developed on the basis of the model
of Ref.  \cite{Drechsel}. One of the modifications 
made in Ref. \cite{Azn0} 
consisted in the incorporation of Regge poles with increasing energies.
This allowed us to describe
pion photoproduction multipole amplitudes \cite{GWU0,GWU3}
with a unified Breit-Wigner parametrization of 
resonance contributions in the form close to that 
introduced by Walker \cite{Walker}.
The Regge-pole amplitudes were constructed
using a gauge invariant Regge-trajectory-exchange
model developed in  Refs. \cite{Laget1,Laget2}. This model
gives a good description of the pion photoproduction data above
the resonance region and can be extended to
finite $Q^2$  \cite{Laget3}. 

The incorporation of Regge poles into the 
background of UIM, built from
the nucleon exchanges in the $s$-
and $u$-channels and $t$-channel
$\pi$, $\rho$ and $\omega$ exchanges,
was made in Ref. \cite{Azn0} in the following way:
\begin{eqnarray} 
&&Background\\
&&=[N+\pi+\rho+\omega]_{UIM}~at~s<s_0,\nonumber\\ 
&&=[N+\pi+\rho+\omega]_{UIM}\frac{1}{1+(s-s_0)^2}+\nonumber\\
&&Re[\pi+\rho+\omega+b_1+a_2]_{Regge}
\frac{(s-s_0)^2}{1+(s-s_0)^2}~at~s>s_0.\nonumber 
\end{eqnarray}
Here the Regge-pole amplitudes were taken from
Refs. \cite{Laget1,Laget2} and consisted of
reggeized $\pi$, $\rho$, $\omega$,  $b_1$, 
and $a_2$ $t$-channel exchange
contributions.
This background was unitarized in the $K$-matrix approximation.
The value of $s_0\simeq 1.2~$GeV$^2$ 
was found in Ref. \cite{Azn0}
from the description of the 
pion photoproduction multipole amplitudes \cite{GWU0,GWU3}.
With this value of $s_0$, we obtained a good description of
$\pi$ electroproduction data
at $Q^2=0.4$ and $0.65~$GeV$^2$ 
in the first, second and 
third resonance regions \cite{Azn04,Azn065}.
The modification of Eq. (8) was important
to obtain a better description of 
the data in the second and third 
resonance regions,
but played an insignificant role at $\sqrt{s}<1.4~$GeV.

When the relation in Eq. (8) was applied
for $Q^2 \geq 0.9~$GeV$^2$, 
the best description of the data was obtained with 
$\sqrt s_0 > 1.8~$GeV. 
Consequently, in the analysis of the data \cite{Park},
the background of UIM was built just from 
the nucleon exchanges in the $s$-
and $u$-channels and $t$-channel
$\pi$, $\rho$ and $\omega$ exchanges.

\section{$N,\pi,\rho,\omega$ contributions}

In both approaches, DR and UIM,
the non-resonant background contains Born terms
corresponding to
the $s$- and $u$-channel nucleon exchanges
and $t$-channel pion contribution,
and therefore depends 
on the proton, neutron,
and pion form factors.
The background of the UIM also contains  
the $\rho$ and $\omega$ $t$-channel exchanges 
and, therefore, the contribution
of the form factors
$G_{\rho(\omega)\rightarrow\pi\gamma}(Q^2)$.
All these form factors, except the neutron
electric and $G_{\rho(\omega)\rightarrow\pi\gamma}(Q^2)$ ones,
are known in the region of $Q^2$
that is the subject of this study.
For the proton form factors we used the parametrizations
found for the existing data 
in Ref. \cite{Melnitchouk}. The neutron magnetic form factor
and the pion form factor
were taken from Refs. \cite{Lung,Lachniet} and 
\cite{Bebek1,Bebek2,Horn,Tadevos},
respectively. 
The neutron electric form factor, $G_{E_n}(Q^2)$, is measured
up to $Q^2=1.45~$GeV$^2$ \cite{Madey}, and Ref. \cite{Madey}
presents a parametrization 
for all existing data on  $G_{E_n}(Q^2)$,
which we used 
for the extrapolation of
$G_{E_n}(Q^2)$ to  $Q^2>1.45~$GeV$^2$. 
In our final results at high $Q^2$, we allow for up to a
$50\%$ deviation from this parametrization that is accounted
for in the systematic uncertainty. 
There are no measurements of the form factors
$G_{\rho(\omega)\rightarrow\pi\gamma}(Q^2)$;
however, investigations made using 
both QCD sum rules \cite{Eletski}
and a quark model \cite{AznOgan} predict a $Q^2$ dependence
of $G_{\rho(\omega)\rightarrow\pi\gamma}(Q^2)$
close to the dipole form 
$G_D(Q^2)
=1/(1+\frac{Q^2}{0.71GeV^2})^2$.
We used this dipole form in our analysis
and introduced in our final results a systematic
uncertainty that accounts for a $20\%$ deviation
from $0.71~$GeV$^2$.
All uncertainties, including those arising
from the measured proton, neutron and pion form factors,
were added in quadrature and will be,
as one part of our total model uncertainties, referenced as
model uncertainties (I) of our results.

\section{Resonance contributions}

We have taken into account
all well-established 
resonances from the first, second, and third resonance
regions. These are 4- and 3-star resonances:
$\Delta(1232)P_{33}$, $N(1440)P_{11}$,
$N(1520)D_{13}$, $N(1535)S_{11}$,
$\Delta(1600)P_{33}$, $\Delta(1620)S_{31}$,
$N(1650)S_{11}$,
$N(1675)D_{15}$,
$N(1680)F_{15}$,
$N(1700)D_{13}$, $\Delta(1700)D_{33}$,
$N(1710)P_{11}$, and
$N(1720)P_{13}$.
For the masses, widths, and  $\pi N$ branching
ratios  of these resonances we used
the mean values of the data from the Review of Particle
Physics (RPP) \cite{PDG}. They are presented in 
Table \ref{parameters}.
Resonances of the fourth
resonance region have no influence
in the energy region under investigation
and were not included.

Resonance contributions to the multipole amplitudes
were parametrized in the usual Breit-Wigner form
with energy-dependent widths \cite{Walker}. 
An exception
was made  for
the $\Delta(1232)P_{33}$ resonance, which was treated
differently. According to the phase-shift analyses
of $\pi N$ scattering,
the $\pi N$ amplitude corresponding to
the $\Delta(1232)P_{33}$ resonance is elastic
up to $W=1.43~$GeV (see, for example, the
latest GWU analyses \cite{GWU1,GWU2}).
In combination with DR and Watson's theorem, this provides
strict constraints on the multipole amplitudes
$M_{1+}^{3/2}$, $E_{1+}^{3/2}$, $S_{1+}^{3/2}$
that correspond to the 
$\Delta(1232)P_{33}$ resonance \cite{Azn0}.
In particular, it was shown \cite{Azn0} 
that with increasing $Q^2$, 
the $W$-dependence of $M_{1+}^{3/2}$
remains unchanged
and close to that 
from the GWU analysis \cite{GWU3} at $Q^2=0$,
if the same normalizations of the amplitudes
at the resonance position are used.
This constraint on the large 
$M_{1+}^{3/2}$ amplitude plays an 
important role in the reliable
extraction of the amplitudes for  the
$\gamma^* N\rightarrow \Delta(1232)P_{33}$ transition.
It also impacts the analysis
of the second resonance region,
because resonances from this
region overlap with the $\Delta(1232)P_{33}$.

\begin{table*}
\begin{center}
\begin{tabular}{cccccccccccccccccccccccc}
\hline
&&&&&&&&&&\\
$N^*$&&&&$M$(MeV)&&&$\tilde{M}$(MeV)&&&$\Gamma$(MeV)&&&
$\tilde{\Gamma}$(MeV)&&&$\beta_{\pi N}(\%)$&&&
$\tilde{\beta}_{\pi N}(\%)$&\\
&&&&&&&&&&\\
\hline
&&&&&&&&&&\\
$\Delta(1232)P_{33}$&&&&$1231-1233$&&&$1232$&&&$116-120$&&&  
$118$&&&$100$&&&$100$&\\
&&&&&&&&&&\\
$N(1440)P_{11}$&&&&$1420-1470$&&&$1440$&&&$200-450$&&&
$350$&&&$55-75$&&&$60$&\\
&&&&&&&&&&\\
$N(1520)D_{13}$&&&&$1515-1525$&&&$1520$&&&$100-125$&&&
$112$&&&$55-65$&&&$60$&\\
&&&&&&&&&&\\
$N(1535)S_{11}$&&&&$1525-1545$&&&$1535$&&&$125-175$&&&
$150$&&&$35-55$&&&$45$&\\
&&&&&&&&&&\\
$\Delta(1600)P_{33}$&&&&$1550-1700$&&&$1600$&&&$250-450$&&&  
$350$&&&$10-25$&&&$20$&\\
&&&&&&&&&&\\
$\Delta(1620)S_{31}$&&&&$1600-1660$&&&$1630$&&&$135-150$&&&  
$145$&&&$20-30$&&&$25$&\\
&&&&&&&&&&\\
$N(1650)S_{11}$&&&&$1645-1670$&&&$1655$&&&$145-185$&&&
$165$&&&$60-95$&&&$75$&\\
&&&&&&&&&&\\
$N(1675)D_{15}$&&&&$1670-1680$&&&$1675$&&&$130-165$&&&
$150$&&&$35-45$&&&$40$&\\
&&&&&&&&&&\\
$N(1680)F_{15}$&&&&$1680-1690$&&&$1685$&&&$120-140$&&&
$130$&&&$65-70$&&&$65$&\\
&&&&&&&&&&\\
$N(1700)D_{13}$&&&&$1650-1750$&&&$1700$&&&$50-150$&&&
$100$&&&$5-15$&&&$10$&\\
&&&&&&&&&&\\ 
$\Delta(1700)D_{33}$&&&&$1670-1750$&&&$1700$&&&$200-400$&&&
$300$&&&$10-20$&&&$15$&\\
&&&&&&&&&&\\
$N(1710)P_{11}$&&&&$1680-1740$&&&$1710$&&&$50-250$&&&
$100$&&&$10-20$&&&$15$&\\
&&&&&&&&&&\\
$N(1720)P_{13}$&&&&$1700-1750$&&&$1720$&&&$150-300$&&&   
$200$&&&$10-20$&&&$15$&\\
&&&&&&&&&&\\
\hline
\end{tabular}
\caption{\label{parameters} List of masses, widths, and branching
ratios of the resonances included in our analysis.
The quoted ranges are taken from RPP \cite{PDG}.
The quantities labeled by tildes 
($\tilde{M}$,
$\tilde{\Gamma}$, 
$\tilde{\beta}_{\pi N}$)
correspond to the values
used in the analysis and in the extraction
of the $\gamma^* p\rightarrow N^*$
helicity amplitudes.}
\end{center}
\end{table*} 

The fitting parameters in our analyses were 
the $\gamma^* p\rightarrow N^{*}$ helicity amplitudes,
$A_{1/2}$, $A_{3/2}$, $S_{1/2}$.
They are related to the resonant portions of the multipole
amplitudes at the resonance positions.
For the resonances with
$J^P=\frac{1}{2}^{-},\frac{3}{2}^{+},...$, these relations are
the following: 
\begin{eqnarray}
&&A_{1/2}=-\frac{1}{2}\left[(l+2){\cal E}_{l+}
+l{\cal M}_{l+}\right], \\
&&A_{3/2}=\frac{\left[l(l+2)\right]^{1/2}}{2}
({\cal E}_{l+}-{\cal M}_{l+}),\\
&&S_{1/2}=-\frac{1}{\sqrt{2}}(l+1){\cal S}_{l+}.
\end{eqnarray}
For the resonances with
$J^P=\frac{1}{2}^{+},\frac{3}{2}^{-},...$:
\begin{eqnarray}
&&A_{1/2}=\frac{1}{2}\left[
(l+2){\cal M}_{(l+1)-}-l{\cal E}_{(l+1)-}\right],\\
&&A_{3/2}=-\frac{\left[l(l+2)\right]^{1/2}}{2}
({\cal E}_{(l+1)-}+{\cal M}_{(l+1)-}),\\
&&S_{1/2}=-\frac{1}{\sqrt{2}}(l+1)
{\cal S}_{(l+1)-},
\end{eqnarray}
where $J$ and $P$ are the  
spin and parity of the 
resonance, $l=J-\frac{1}{2}$, and

\begin{eqnarray}
&{\cal M}_{l\pm}({\cal E}_{l\pm},{\cal S}_{l\pm})\equiv 
aImM^R_{l\pm}(E^R_{l\pm},S^R_{l\pm})(W=M),
\\
&a\equiv\frac{1}{C_I}
\left[(2J+1)\pi\frac{q_{r}}{K}\frac{M}{m}
\frac{\Gamma}{\beta_{\pi N}}\right]^{1/2},\nonumber\\
&C_{1/2}=-\sqrt{\frac{1}{3}},~C_{3/2}=\sqrt{\frac{2}{3}}
~for~\gamma^* p\rightarrow \pi^0 p,\nonumber\\
&C_{1/2}=-\sqrt{\frac{2}{3}},~C_{3/2}=-\sqrt{\frac{1}{3}}
~for~\gamma^* p\rightarrow \pi^+ n.\nonumber
\end{eqnarray}
Here $C_I$ are the isospin
Clebsch-Gordon coefficients in the decay $N^*\rightarrow \pi N$;
$\Gamma$, $M$, and $I$ are the total width, mass, 
and isospin of the 
resonance, respectively, $\beta_{\pi N}$ is its branching ratio
to the $\pi N$ channel,
$K$ and $q_r$ are the photon equivalent energy and 
the pion momentum
at the resonance position in c.m. system.
For the transverse amplitudes $A_{1/2}$ and $A_{3/2}$,
these relations were introduced by Walker \cite{Walker};
for the longitudinal amplitudes, they agree with those 
from Refs. \cite{Arndt,Capstick,Kamalov1}.

The masses, widths, and  $\pi N$ branching
ratios  of the resonances
are known in the ranges  presented
in Table \ref{parameters}.
The uncertainties of masses and widths
of the  $N(1440)P_{11}$,
$N(1520)D_{13}$, and $N(1535)S_{11}$ are quite significant
and can affect the
resonant portions of the multipole
amplitudes for these resonances at the resonance positions.
These uncertainties
were taken into account by refitting the data multiple times with
the width (mass) of each of the resonances changed within one
standard deviation\footnote{The standard deviations were 
defined as
$\sigma_M= (M_{max}-M_{min})/ \sqrt{12}$ and $\sigma_\Gamma =
(\Gamma_{max}-\Gamma_{min})/ \sqrt{12}$, with the maximum
and minimum values as shown in Table V.}  while keeping those for
other resonances fixed. The resulting uncertainties
of the $\gamma^* p\rightarrow
N(1440)P_{11}$,
$N(1520)D_{13}$, $N(1535)S_{11}$ amplitudes
were added in quadrature
and considered as model
uncertainties (II).

In Sec. II, we discussed
that in the analysis of the data
reported in Table \ref{pip_data},
there is another uncertainty in the amplitudes
for the $N(1440)P_{11}$,
$N(1520)D_{13}$, and $N(1535)S_{11}$,
which is caused by the limited information available
on magnitudes of resonant amplitudes in the
third resonance region.
To evaluate  the influence of these states
on the extracted
$\gamma^*p\rightarrow N(1440)P_{11}$,
$N(1520)D_{13}$, $N(1535)S_{11}$ amplitudes,
we used two ways of estimating
their strength.

(i) Directly including these states in the fit,
taking the corresponding amplitudes $A_{1/2}$, $A_{3/2}$,
$S_{1/2}$ as free parameters.

(ii) Applying some constraints on their amplitudes.
Using symmetry relations within the  $[70,1^-]$ multiplet
given by the single quark transition model  \cite{SQTM},
we have related
the transverse amplitudes for
the members of this multiplet
($\Delta(1620)S_{31}$,
$N(1650)S_{11}$,
$N(1675)D_{15}$, $N(1700)D_{13}$, and $\Delta(1700)D_{33}$)
to the amplitudes
of $N(1520)D_{13}$ and $N(1535)S_{11}$ that are well
determined in the analysis.
The longitudinal amplitudes of these resonances
and the amplitudes of the resonances
$\Delta(1600)P_{33}$ and
$N(1710)P_{11}$, which have small photocouplings \cite{PDG}
and are not seen in low $Q^2$ $\pi$ and 2$\pi$
electroproduction \cite{Azn065},
were assumed to be zero.

The results obtained
for $N(1440)P_{11}$,
$N(1520)D_{13}$, and $N(1535)S_{11}$ using the two procedures
are very close to each other. The amplitudes
for these resonances
presented below are the average values of the results
obtained in these fits. 
The uncertainties arising from this
averaging procedure 
were added in quadrature 
to the model uncertainties (II).

\section{Results}

Results for the extracted 
$\gamma^*p\rightarrow \Delta(1232)P_{33}$, $N(1440)P_{11}$,
$N(1520)D_{13}$, $N(1535)S_{11}$ amplitudes
are presented in Tables \ref{m1+}-\ref{D13_av}.
Here we show separately the amplitudes  
obtained in the DR and UIM approaches.
The amplitudes are presented with the fit errors and 
model uncertainties caused
by the $N,\pi,\rho$, and $\omega$
contributions to the background,
and those caused by the masses and widths
of the  $N(1440)P_{11}$,
$N(1520)D_{13}$, and $N(1535)S_{11}$,
and by the resonances of the third resonance region. 
These uncertainties, discussed in Sections 
IV and V, and  referred to as model
uncertainties (I) and (II), were added in quadrature
and represent model uncertainties  of the DR and UIM results. 

The DR and UIM approaches give comparable descriptions
of the data (see $\chi^2$ values in Tables 
\ref{pol_data}-\ref{pi0_data}), and, therefore,
the differences in $A_{1/2},A_{3/2},S_{1/2}$ 
are related only to the model assumptions.
We, therefore, ascribe the difference in  the results 
obtained in the two approaches to model uncertainty,
and present as our final results  
in Tables \ref{m1+}-\ref{S11} and \ref{D13_av} the mean 
values of the amplitudes extracted using DR and UIM.
The uncertainty that originates from the averaging
is considered as an additional model uncertainty - uncertainty (III). 
Along with the average values of the uncertainties (I) and (II)
obtained in the DR and UIM approaches, it is 
included
in quadrature in the total model uncertainties of the average
amplitudes.

In the fit we have included the experimental
point-to-point systematics by adding them in quadrature
with the statistical error.
We also took into account
the overall normalization error of the CLAS cross sections
data which is about 5\%.  It was
checked that the overall normalization error
results in modifications of all extracted
amplitudes, except $M_{1+}^{3/2}$, that are significantly
smaller than the fit errors of these amplitudes.
For $M_{1+}^{3/2}$, this error
results in the overall normalization error 
which is larger
than the fit error. It is about 2.5\% for low $Q^2$, and increases
up to 3.2-3.3\% at $Q^2=3-6~$GeV$^2$. For $M_{1+}^{3/2}$, 
the fit error given in Table \ref{m1+} includes
the overall normalization error added in quadrature to the fit error.

\begin{figure*}[htp]
\begin{center}
\includegraphics[width=12.0cm]{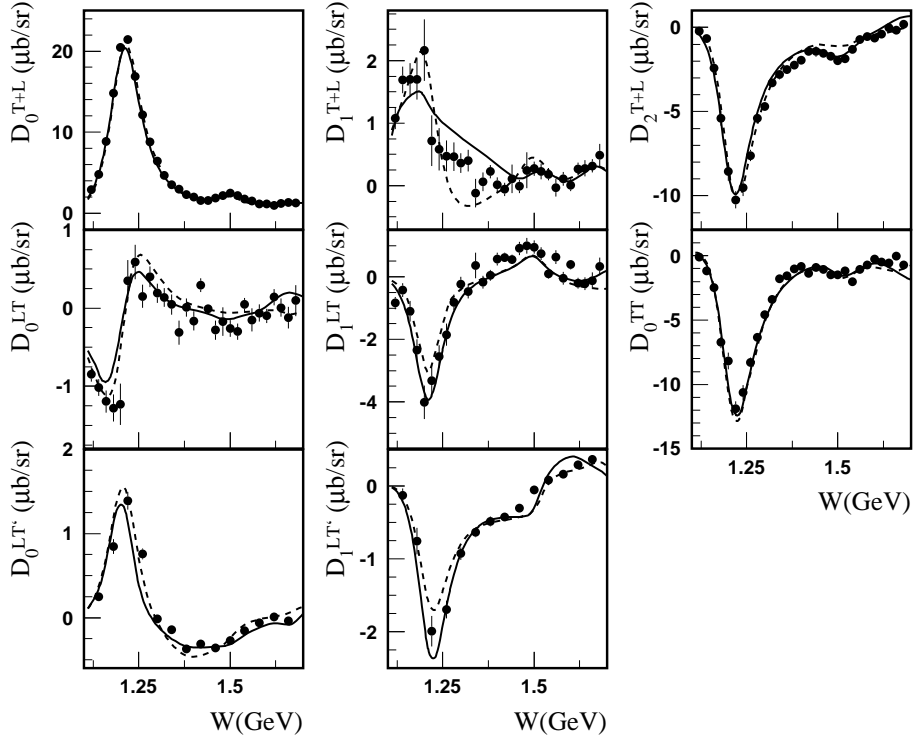}
\vspace{-0.1cm}
\caption{\small 
Our results for the Legendre moments of the
$\vec{e}p\rightarrow ep\pi^0$ structure functions
in comparison with experimental data \cite{Joo1} for $Q^2=0.4~$GeV$^2$.
The solid (dashed) curves
correspond to the results obtained using DR (UIM) approach.
\label{leg_pi0_04}}
\end{center}
\end{figure*}

\begin{figure*}[htp]
\begin{center}
\includegraphics[width=12.0cm]{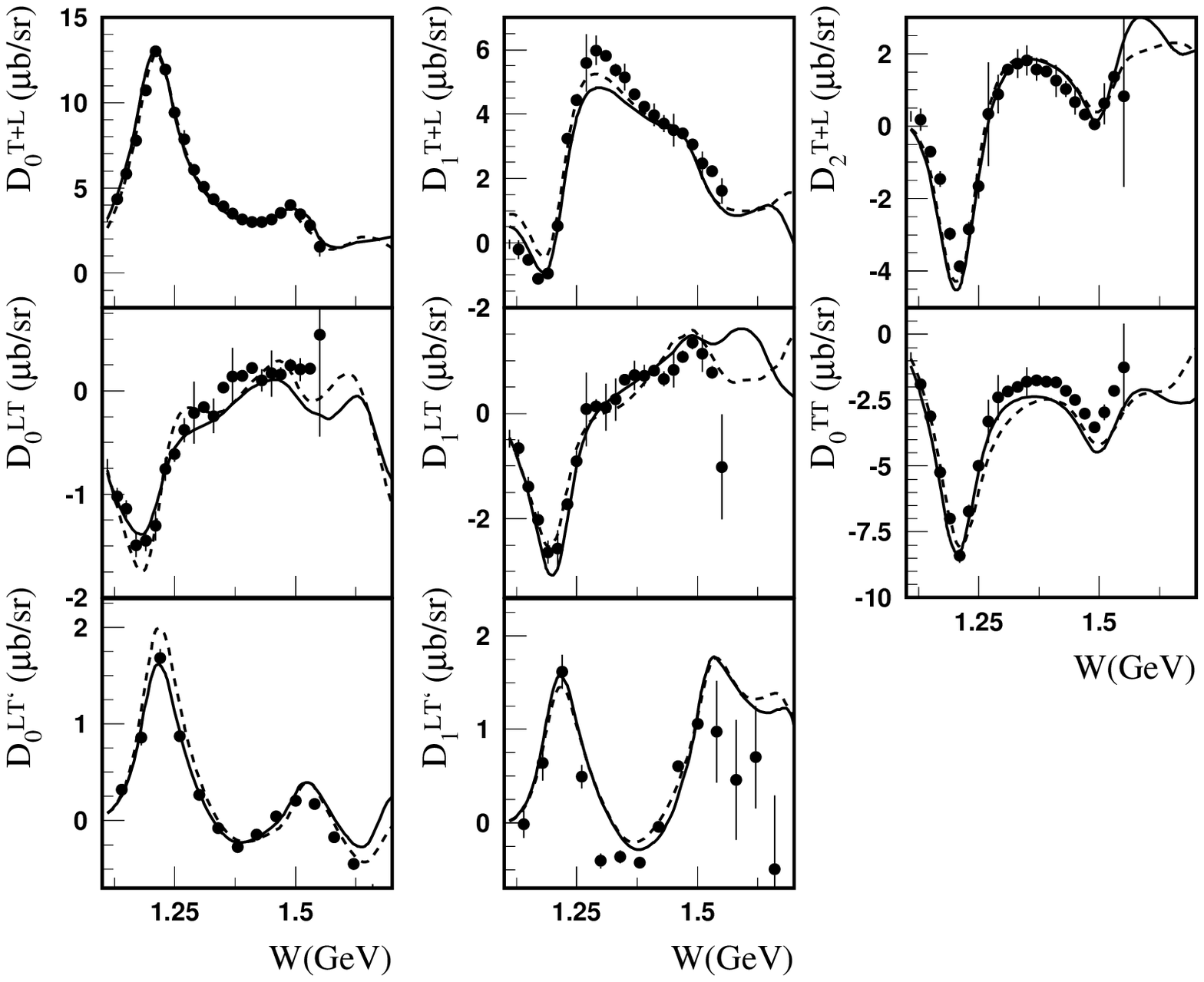}
\vspace{-0.1cm}
\caption{\small 
Our results for the Legendre moments of the
$\vec{e}p\rightarrow en\pi^+$ structure functions
in comparison with experimental data \cite{Egiyan} for 
$Q^2=0.4~$GeV$^2$.
The solid (dashed) curves
correspond to the results obtained using DR (UIM) approach.
\label{leg_pip_04}}
\end{center}
\end{figure*}

\begin{figure*}[htp]
\begin{center}
\includegraphics[width=12.0cm]{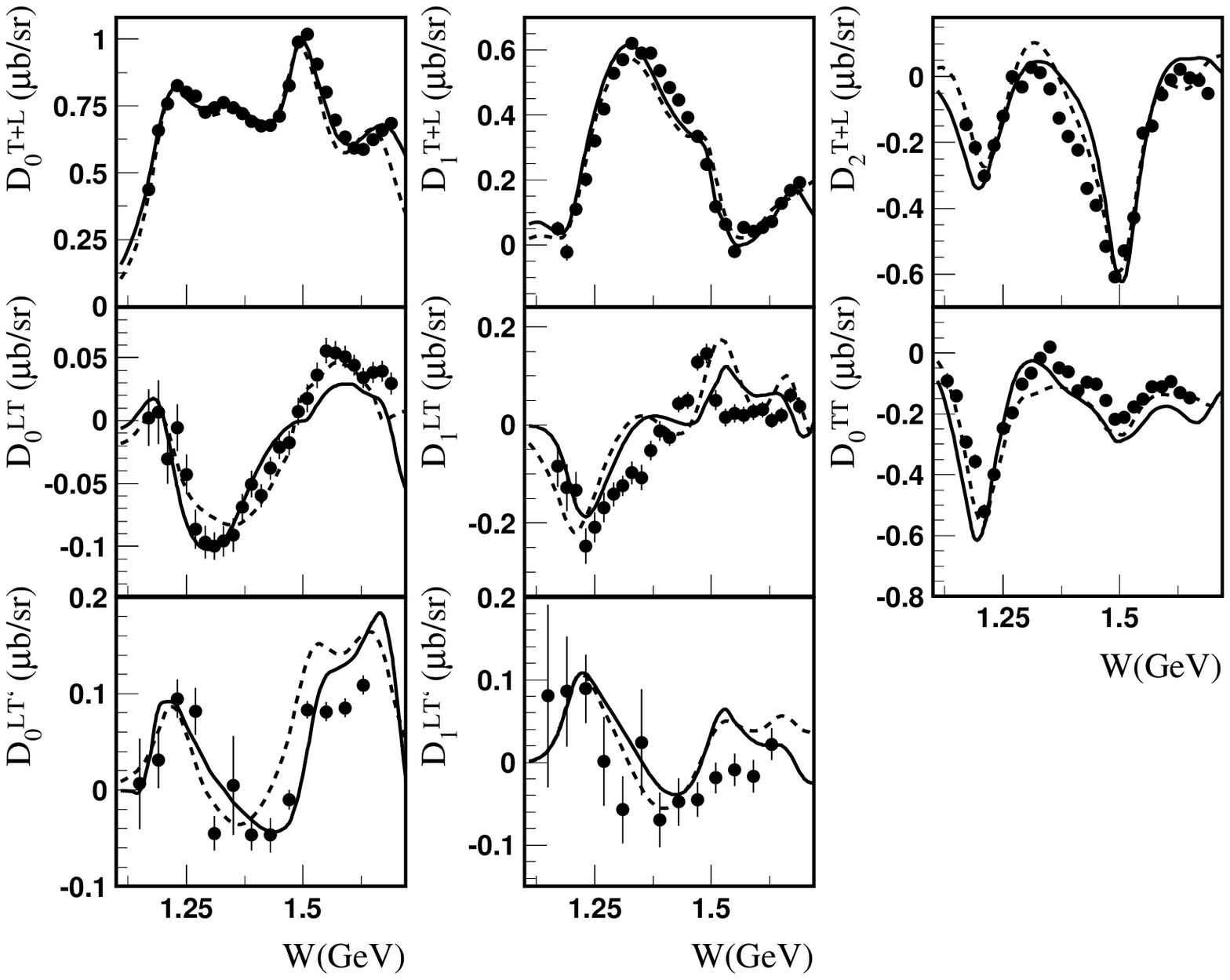}
\vspace{-0.1cm}
\caption{\small 
Our results for the Legendre moments of the
$\vec{e}p\rightarrow en\pi^+$ structure functions
in comparison with experimental data \cite{Park} for $Q^2=2.44~$GeV$^2$.
The solid (dashed) curves
correspond to the results obtained using DR (UIM) approach.
\label{leg_244}}
\end{center}
\end{figure*}

\begin{figure*}[htp]
\begin{center}
\includegraphics[width=12.0cm]{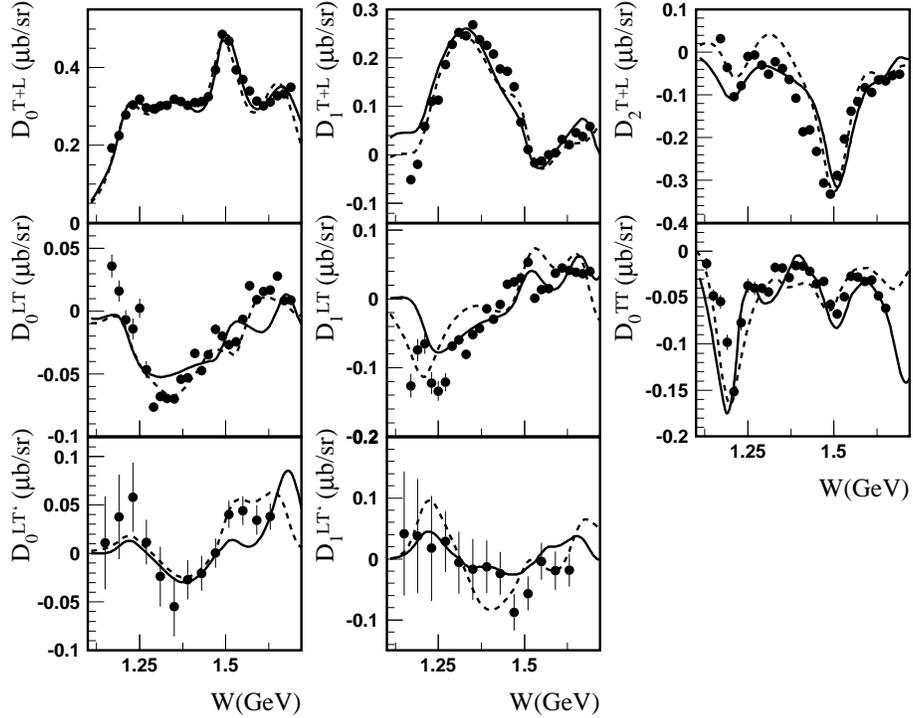}
\vspace{-0.1cm}
\caption{\small 
The same as in Fig. \ref{leg_244} for $Q^2=3.48~$GeV$^2$.
\label{leg_348}}
\end{center}
\end{figure*}

\begin{figure*}[htp] \begin{center} 
\includegraphics[width=7.2cm]{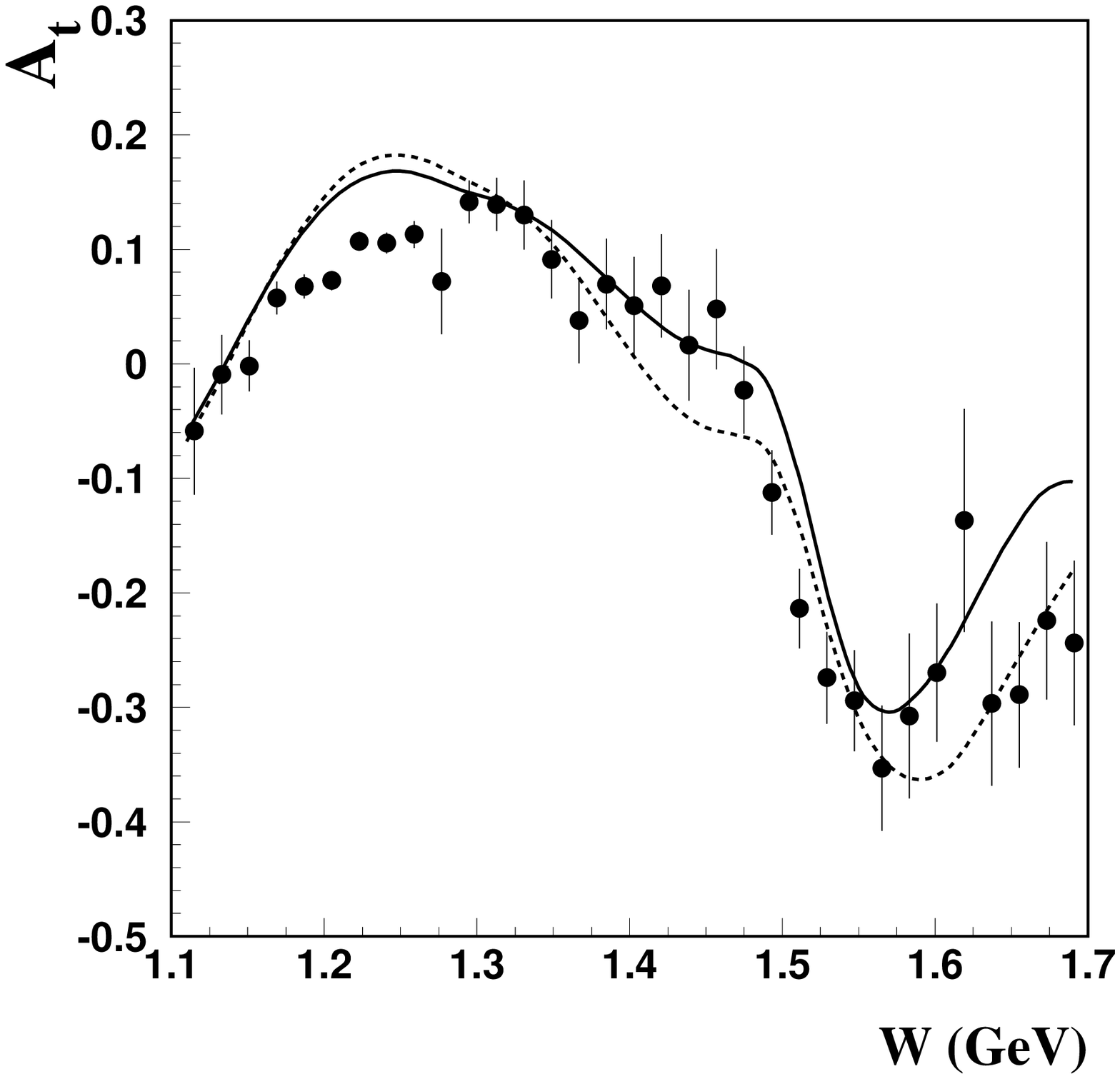} 
\includegraphics[width=7.2cm]{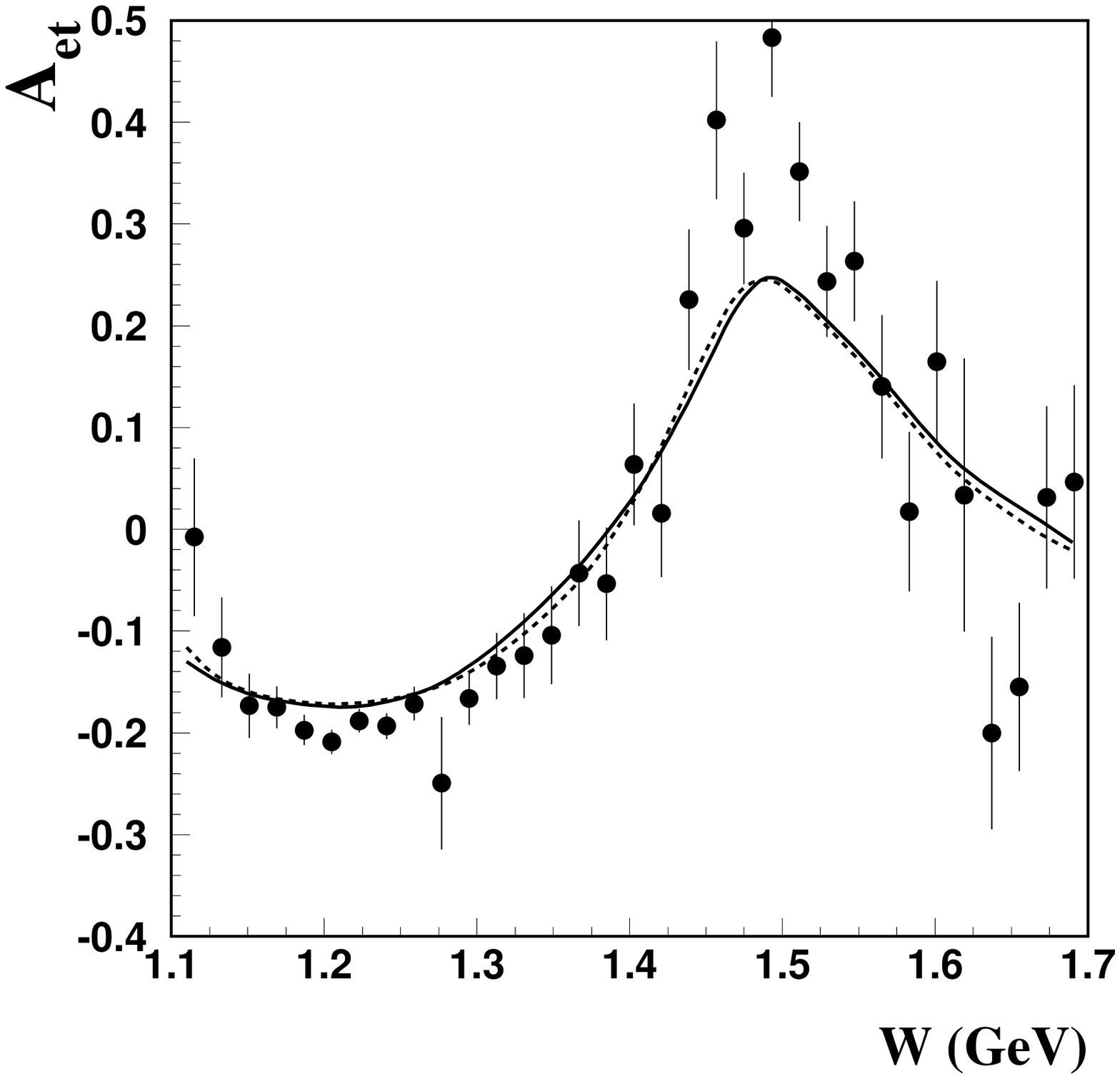}
\vspace{-0.1cm} \caption{\small 
$A_{t}$ (left panel) and $A_{et}$ (right panel) as functions of the 
invariant
mass $W$, integrated over the whole range in $\cos{\theta}$,
$0.252<Q^2<0.611~$GeV$^2$ and $60^0<\phi<156^0$.
Experimental data are form Ref. \cite{Biselli}.
Solid and dashed curves correspond to our results
obtained using DR and UIM approaches,
respectively.  
\label{at}} 
\end{center} 
\end{figure*}

\begin{figure*}[ht]
\includegraphics[width=12cm]{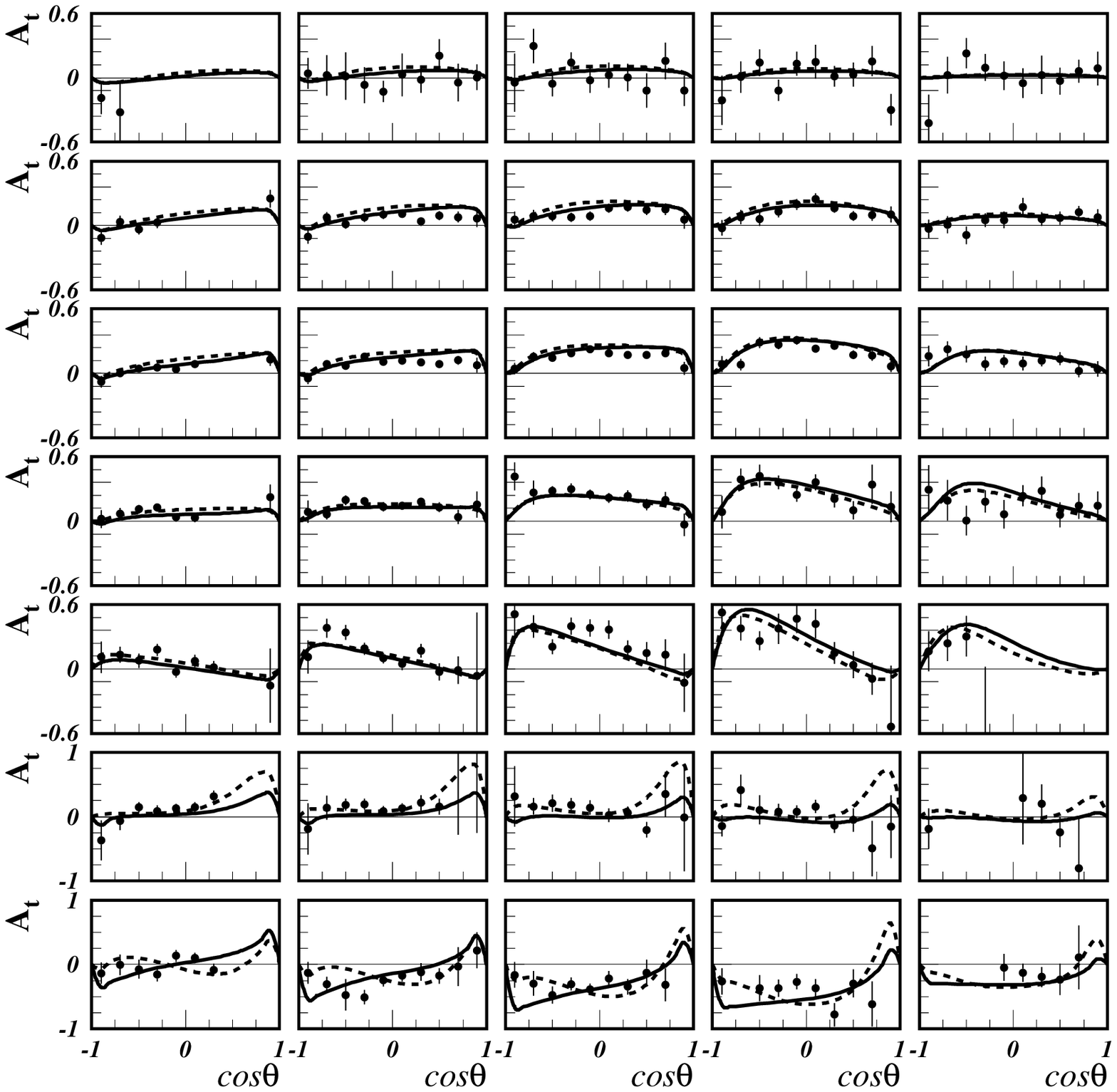}
\vspace{-0.1cm}
\caption{\small 
Our results for the longitudinal target asymmetry $A_t$
in comparison with experimental data for $Q^2=0.385~$GeV$^2$ \cite{Biselli}.
Solid (dashed) curves
correspond to the results obtained using DR (UIM) approach.
Rows correspond to 7 $W$ bins with $W$ mean values of 1.125, 1.175,
1.225, 1.275, 1.35, 1.45, and 1.55 GeV.
Columns correspond to $\phi$ bins with
$\phi=\pm 72^0,\pm 96^0,\pm 120^0,\pm 144^0,\pm 168^0$.
The solid circles are the average values of the data for positive 
$\phi$'s 
and those
at negative  $\phi$'s taken with opposite
signs.
\label{at_04}}
\end{figure*}

\begin{figure*}[ht]
\includegraphics[width=12cm]{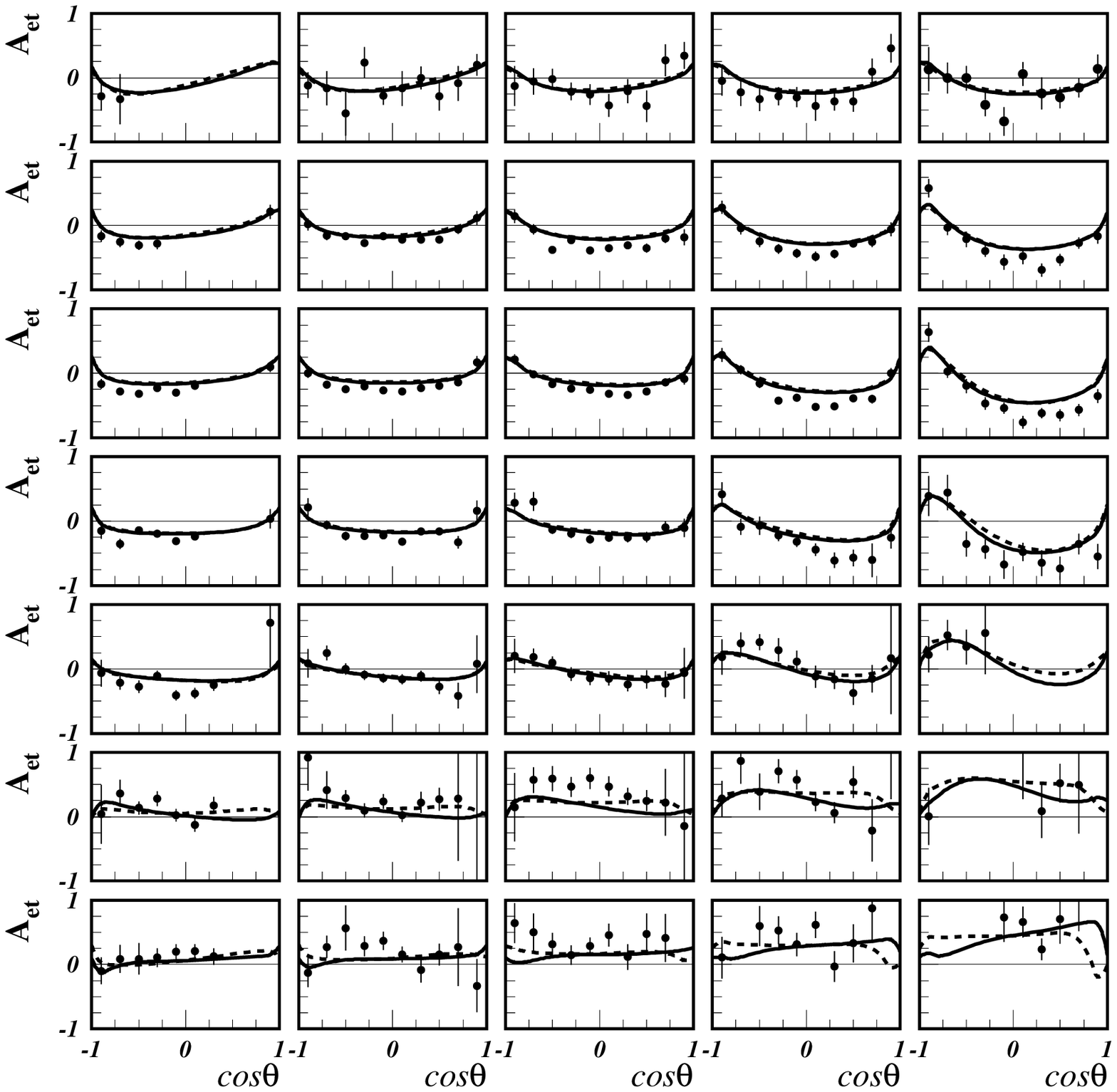}
\vspace{-0.1cm}
\caption{\small 
Our results for the beam-target asymmetry $A_{et}$
in comparison with experimental data for $Q^2=0.385~$GeV$^2$ \cite{Biselli}.
Solid (dashed) curves
correspond to the results obtained using DR (UIM) approach.
Rows correspond to 7 $W$ bins with $W$ mean values of 1.125, 1.175,
1.225, 1.275, 1.35, 1.45, and 1.55 GeV.
Columns correspond to $\phi$ bins with
$\phi=\pm 72^0,\pm 96^0,\pm 120^0,\pm 144^0,\pm 168^0$.
The average values of the data for positive and negative $\phi$'s
are shown by solid circles.
\label{aet_04}}
\end{figure*}

Examples of the comparison with the experimental data
are presented 
in Figs. \ref{leg_pi0_04}-\ref{str_3}. 
The obtained values of $\chi^2$ in the fit 
to the data are presented
in Tables \ref{pol_data}-\ref{pi0_data}.
The relatively large values of $\chi^2$ for 
$\frac{d\sigma}{d\Omega}(\pi^0)$ at
$Q^2= 0.16,0.2~$GeV$^2$ and for 
$\frac{d\sigma}{d\Omega}(\pi^+)$ at
$Q^2= 0.3,0.4~$GeV$^2$ and 
$Q^2= 1.72,2.05~$GeV$^2$ 
are caused by small statistical errors,
which for each data set \cite{Smith}, \cite{Egiyan}
and \cite{Park}, increase with increasing $Q^2$.
The values of $\chi^2$ for $A_{LT'}$ 
at $Q^2\ge 1.72~$GeV$^2$ are somewhat large.
However, as demonstrated in 
Figs. \ref{leg_244},\ref{leg_348}, the  description 
on the whole is satisfactory.
 
\begin{table*}[t]
\begin{tabular}{|l|cc|}
\hline
&&\\
$~~Q^2$&$~~~~~~~~~~~~~~~~~~~~~~~~~~~
ImM^{3/2}_{1+}$($\sqrt{\mu b}$), W=1.232 GeV&\\
(GeV$^2$)&&\\
&DR$~~~~~~~~~~~~~~~~~~~~~~~~~~~~~~~~$UIM$~~~$& Final results\\
&&\\
\hline
0.3&$~~~5.173\pm 0.130~~~~~~~~~~~~~~5.122\pm 0.130\pm 0.004$
&$~~~~~~5.148\pm 0.130\pm 0.026~$\\
0.4&$~~~4.843\pm 0.122~~~~~~~~~~~~~~4.803\pm 0.122\pm 0.005$
&$~~~~~~4.823\pm 0.122\pm 0.021~$\\
0.525&$~~~4.277\pm 0.109~~~~~~~~~~~~~~4.238\pm 109\pm 0.008$
&$~~~~~~4.257\pm 0.109\pm 0.021~$\\
0.65&$~~~3.814\pm 0.097~~~~~~~~~~~~~~3.794\pm 0.097\pm 0.009$
&$~~~~~~3.804\pm 0.097\pm 0.013~$\\
0.75&$~~~3.395\pm 0.088~~~~~~~~~~~~~~3.356\pm 0.088\pm 0.011$
&$~~~~~~3.375\pm 0.088\pm 0.022~$\\
0.9&$~~~3.010\pm 0.078~~~~~~~~~~~~~~2.962\pm 0.078\pm 0.012$
&$~~~~~~2.986\pm 0.078\pm 0.027~$\\
1.15&$~~~2.487\pm 0.066~~~~~~~~~~~~~~2.438\pm 0.066\pm 0.013$
&$~~~~~~2.463\pm 0.066\pm 0.028~$\\
1.45&$~~~1.948\pm 0.059~~~~~~~~~~~~~~1.880\pm 0.059\pm 0.014$
&$~~~~~~1.914\pm 0.059\pm 0.037~$\\
3.0&$~~~0.725\pm 0.022\pm 0.011~~~~0.693\pm 0.022\pm 0.016$
&$~~~~~~0.709\pm 0.022\pm 0.023~$\\
3.5&$~~~0.582\pm 0.018\pm 0.012~~~~0.558\pm 0.018\pm 0.017$
&$~~~~~~0.570\pm 0.018\pm 0.021~$\\
4.2&$~~~0.434\pm 0.014\pm 0.014~~~~0.412\pm 0.014\pm 0.018$
&$~~~~~~0.423\pm 0.014\pm 0.021~$\\
5.0&$~~~0.323\pm 0.012\pm 0.021~~~~0.312\pm 0.012\pm 0.023$
&$~~~~~~0.317\pm 0.012\pm 0.024~$\\
6.0&$~~~0.200\pm 0.012\pm 0.024~~~~0.191\pm 0.012\pm 0.027$
&$~~~~~~0.196\pm 0.012\pm 0.027~$\\
\hline
\end{tabular}
\caption{\label{m1+}
The
results for the imaginary part of $M^{3/2}_{1+}$
at $W=1.232~$GeV.
For the DR and UIM results, the first and second uncertainties are
the statistical uncertainty from the fit and 
the model uncertainty (I) (see Sec. IV), respectively.
For $Q^2=0.3-1.45~$GeV$^2$, the uncertainty (I)
is practically related only to 
the form factors 
$G_{\rho,\omega}(Q^2)$;
for this reason it does not affect the
amplitudes found using DR. 
Final results are the average values of the amplitudes
found using DR and UIM; here the first 
uncertainty is statistical, and the second one is the model
uncertainty discussed in Sec. VI. 
 }
\end{table*}
 
\begin{table*}[t]
\begin{tabular}{|l|cc|}
\hline
&&\\
$~~Q^2$&$~~~~~~~~~~~~~~~~~~~~~~~~~~~~R_{EM}$($\%$)&\\
(GeV$^2$)&&\\
&$~~~~~~~$DR$~~~~~~~~~~~~~~~~~~~~~~$UIM$~~~~~~~$&Final results\\
&&\\
\hline
0.16&$~~-2.0\pm 0.1~~~~~~~~~~~-1.7\pm 0.1\pm 0.04$
&$~~~-1.9\pm 0.1\pm 0.2~~$\\
0.2&$~~-1.9\pm 0.2~~~~~~~~~~~-1.6\pm 0.2\pm 0.04$
&$~~~-1.8\pm 0.2\pm 0.2~~$\\
0.24&$~~-2.2\pm 0.2~~~~~~~~~~~-2.1\pm 0.2\pm 0.1$
&$~~~-2.2\pm 0.2\pm 0.1 ~~$\\
0.28&$~~-1.9\pm 0.2~~~~~~~~~~~-1.6\pm 0.2\pm 0.1$
&$~~~-1.8\pm 0.2\pm 0.2~~$\\
0.3&$~~-2.2\pm 0.2~~~~~~~~~~~-2.1\pm 0.2\pm 0.1$
&$~~~-2.1\pm 0.2\pm 0.1~~$\\
0.32&$~~-1.9\pm 0.2~~~~~~~~~~~-1.6\pm 0.2\pm 0.1$
&$~~~-1.8\pm 0.2\pm 0.2~~$\\
0.36&$~~-1.8\pm 0.2~~~~~~~~~~~-1.5\pm 0.3\pm 0.1$
&$~~~-1.7\pm 0.3\pm 0.2~~$\\
0.4&$~~-2.9\pm 0.2~~~~~~~~~~~-2.4\pm 0.2\pm 0.1$
&$~~~-2.7\pm 0.2\pm 0.3~~$\\
0.525&$~~-2.3\pm 0.3~~~~~~~~~~~-2.0\pm 0.3\pm 0.1$
&$~~~-2.2\pm 0.3\pm 0.2~~$\\
0.65&$~~-2.0\pm 0.4~~~~~~~~~~~-1.4\pm 0.3\pm 0.1$
&$~~~-1.7\pm 0.4\pm 0.3~~$\\
0.75&$~~-2.2\pm 0.4~~~~~~~~~~~-1.9\pm 0.4\pm 0.1$
&$~~~-2.1\pm 0.4\pm 0.2~~$\\
0.9&$~~-2.4\pm 0.5~~~~~~~~~~~-2.1\pm 0.5\pm 0.2$
&$~~~-2.2\pm 0.5\pm 0.3~~$\\
1.15&$~~-2.0\pm 0.6~~~~~~~~~~~-2.6\pm 0.5\pm 0.2$
&$~~~-2.3\pm 0.6\pm 0.4~~$\\
1.45&$~~-2.4\pm 0.7~~~~~~~~~~~-2.5\pm 0.7\pm 0.2$
&$~~~-2.5\pm 0.7\pm 0.2~~$\\
3.0&$~~-1.6\pm 0.4\pm 0.1~~~~~-2.3\pm 0.4\pm 0.2$
&$~~~-2.0\pm 0.4\pm 0.4~~$\\
3.5&$~~-1.8\pm 0.5\pm 0.2~~~~~-1.1\pm 0.5\pm 0.3$
&$~~~-1.5\pm 0.5\pm 0.5~~$\\
4.2&$~~-2.3\pm 0.8\pm 0.3~~~~~-2.9\pm 0.7\pm 0.4$
&$~~~-2.6\pm 0.8\pm 0.4~~$\\
5.0&$~~-2.2\pm 1.4\pm 0.3~~~~~-3.2\pm 1.5\pm 0.4$
&$~~~-2.7\pm 1.5\pm 0.6~~$\\
6.0&$~~-2.1\pm 2.5\pm 1.1~~~~~-3.6\pm 2.6\pm 1.5$
&$~~~-2.8\pm 2.6\pm 1.7~~$\\
\hline
\end{tabular}
\caption{\label{REM}
The
results for the ratio 
$R_{EM}\equiv ImE^{3/2}_{1+}/ImM^{3/2}_{1+}$
at $W=1.232~$GeV. All other 
relevant information is as given in the
legend of Table \ref{m1+}.
}
\end{table*}
 
\begin{table*}[t]
\begin{tabular}{|l|cc|}
\hline
&&\\
$~~Q^2$&$~~~~~~~~~~~~~~~~~~~~~~~~~~~~R_{SM}$($\%$)&\\
(GeV$^2$)&&\\
&$~~~~~~~$DR$~~~~~~~~~~~~~~~~~~~~~~$UIM$~~~~~~~$&Final results\\
&&\\
\hline
0.16&$~~-4.8\pm 0.2~~~~~~~~~~~~-4.6\pm 0.2\pm 0.04$
& $~~~-4.7\pm 0.2\pm 0.1~~$\\
0.2&$~~-4.9\pm 0.2~~~~~~~~~~~~-4.4\pm 0.2\pm 0.1$
& $~~~-4.7\pm 0.2\pm 0.3~~$\\
0.24&$~~-4.7\pm 0.3~~~~~~~~~~~~-4.5\pm 0.3\pm 0.1$
& $~~~-4.6\pm 0.3\pm 0.1~~$\\
0.28&$~~-5.6\pm 0.3~~~~~~~~~~~~-5.4\pm 0.3\pm 0.1$ 
& $~~~-5.5\pm 0.3\pm 0.1~~$\\
0.3&$~~-5.4\pm 0.2~~~~~~~~~~~~-5.0\pm 0.2\pm 0.1$  
& $~~~-5.2\pm 0.2\pm 0.2~~$\\
0.32&$~~-5.9\pm 0.3~~~~~~~~~~~~-5.5\pm 0.3\pm 0.1$
& $~~~-5.7\pm 0.3\pm 0.2~~$\\
0.36&$~~-5.5\pm 0.3~~~~~~~~~~~~-5.2\pm 0.3\pm 0.1$ 
& $~~~-5.4\pm 0.3\pm 0.2~~$\\
0.4&$~~-5.9\pm 0.2~~~~~~~~~~~~-5.2\pm 0.2\pm 0.1$
& $~~~-5.5\pm 0.2\pm 0.4~~$\\
0.525&$~~-6.0\pm 0.3~~~~~~~~~~~~-5.5\pm 0.3\pm 0.1$
& $~~~-5.8\pm 0.3\pm 0.3~~$\\
0.65&$~~-7.0\pm 0.4~~~~~~~~~~~~-6.2\pm 0.4\pm 0.2$
& $~~~-6.6\pm 0.4\pm 0.4~~$\\
0.75&$~~-7.3\pm 0.4~~~~~~~~~~~~-6.7\pm 0.4\pm 0.2$
& $~~~-7.0\pm 0.4\pm 0.4~~$\\
0.9&$~~-8.6\pm 0.4~~~~~~~~~~~~-8.1\pm 0.4\pm 0.2$
& $~~~-8.4\pm 0.5\pm 0.3~~$\\
1.15&$~~-8.8\pm 0.5~~~~~~~~~~~~-8.0\pm 0.5\pm 0.2$ 
& $~~~-8.4\pm 0.5\pm 0.4~~$\\
1.45&$~~-10.5\pm 0.8~~~~~~~~~~~-9.6\pm 0.8\pm 0.2$ 
& $~~~-10.1\pm 0.8\pm 0.5~~$\\
3.0&$~~-12.6\pm 0.6\pm 0.1~~~~~~~~-11.4\pm 0.6\pm 0.2$
& $~~~-12.0\pm 0.6\pm 0.6~~$\\
3.5&$~~-12.8\pm 0.8\pm 0.3~~~~~~~~-12.4\pm 0.8\pm 0.4$
& $~~~-12.6\pm 0.8\pm 0.4~~$\\
4.2&$~~-17.1\pm 1.2\pm 0.5~~~~~~~~-15.9\pm 1.3\pm 0.7$
& $~~~-16.5\pm 1.3\pm 0.8~~$\\
5.0&$~~-26.6\pm 2.7\pm 1.2~~~~~~~~-25.2\pm 2.7\pm 1.5$
& $~~~-25.9\pm 2.7\pm 1.7~~$\\
6.0&$~~-26.4\pm 5.2\pm 3.2~~~~~~~~-25.3\pm 5.3\pm 3.8$
& $~~~-25.9\pm 5.3\pm 3.8~~$\\
\hline
\end{tabular}
\caption{\label{RSM}
The
results for the ratio 
$R_{SM}\equiv ImS^{3/2}_{1+}/ImM^{3/2}_{1+}$
at $W=1.232~$GeV. All other 
relevant information is as given in the
legend of Table \ref{m1+}.
}
\end{table*}
 
\begin{table*}[t]
\begin{tabular}{|c|ccc|}
\hline
&&&\\
$Q^2$&DR&UIM&Final results\\
(GeV$^2$)&&&\\
&$A_{1/2}~~~~~~~~~~~~~~~~~~~S_{1/2}$
&$A_{1/2}~~~~~~~~~~~~~~~~~~~S_{1/2}$
&$A_{1/2}~~~~~~~~~~~~~~~~~~~S_{1/2}$\\
&&&\\
\hline
0.3&$-15.5\pm 1.2\pm 1.0~~~~31.8\pm 1.8\pm 0.8$&
$~~-24.0\pm 1.2\pm 2.5~~~~37.6\pm 1.9\pm 2.5$
&$~~~~~-19.8\pm 1.2\pm 4.6~~~~34.7\pm 1.8\pm 3.3$\\
0.4&$-9.4\pm 1.1\pm 0.9~~~~30.1\pm 1.4\pm 0.9$
&$~~-19.7\pm 1.1\pm 3.1~~~~34.8\pm 1.3\pm 3.0$
&$~~~~~-14.6\pm 1.1\pm 5.5~~~~32.5\pm 1.3\pm 3.1$\\
0.5&$10.5\pm 1.2\pm 0.9~~~~30.6\pm 1.5\pm 0.9
$&$~~~-4.6\pm 1.3\pm 3.4~~~~36.9\pm 1.6\pm 3.0$
&$~~~~~3.0\pm 1.2\pm 7.9~~~~33.8\pm 1.5\pm 3.7$\\
0.65&$19.5\pm 1.3\pm 1.0~~~~27.6\pm 1.3\pm 1.0$
&$~~~~5.4\pm 1.2\pm 3.4~~~~~~35.2\pm 1.2\pm 3.4$
&$~~~~~12.4\pm 1.2\pm 7.4~~~~31.4\pm 1.2\pm 4.4$\\
0.9&$31.9\pm 2.6\pm 4.3~~~~30.6\pm 2.1\pm 4.3$
&$~~~~~18.7\pm 2.7\pm 4.3~~~~36.2\pm 2.1\pm 4.2$
&$~~~~~25.3\pm 2.7\pm 7.9~~~~33.4\pm 2.1\pm 5.1$\\
1.72&$72.5\pm 1.0\pm 4.4~~~~24.8\pm 1.4\pm 5.4$
&$~~~~~58.5\pm 1.1\pm 4.3~~~~26.9\pm 1.3\pm 5.4$
&$~~~~~65.5\pm 1.0\pm 8.3~~~~25.8\pm 1.3\pm 5.5$\\
2.05&$72.0\pm 0.9\pm 4.3~~~~21.0\pm 1.7\pm 5.1$
&$~~~~~62.9\pm 0.9\pm 3.4~~~~15.5\pm 1.5\pm 5.0$
&$~~~~~67.4\pm 0.9\pm 6.0~~~~18.2\pm 1.6\pm 5.8$\\
2.44&$50.0\pm 1.0\pm 3.4~~~~~9.3\pm 1.3\pm 4.3$
&$~~~~~56.2\pm 0.9\pm 3.4~~~~11.8\pm 1.4\pm 4.3$
&$~~~~~53.1\pm 1.0\pm 4.6~~~~10.6\pm 1.4\pm 4.5$\\
2.91&$37.5\pm 1.1\pm 3.0~~~~~9.8\pm 2.0\pm 2.6$
&$~~~~~42.5\pm 1.1\pm 3.0~~~~13.8\pm 2.1\pm 2.6$
&$~~~~~40.0\pm 1.1\pm 3.9~~~~11.8\pm 2.1\pm 3.3$\\
3.48&$29.6\pm 0.8\pm 2.9~~~~~4.2\pm 2.5\pm 2.6$
&$~~~~~32.6\pm 0.9\pm 2.8~~~~14.1\pm 2.4\pm 2.4$
&$~~~~~31.1\pm 0.9\pm 3.2~~~~9.1\pm 2.5\pm 5.5$\\
4.16&$19.3\pm 2.0\pm 4.0~~~~10.8\pm 2.8\pm 4.7$
&$~~~~~23.1\pm 2.2\pm 4.9~~~~17.5\pm 2.6\pm 5.6$
&$~~~~~21.2\pm 2.1\pm 4.9~~~~14.1\pm 2.7\pm 6.1$\\
\hline
\end{tabular}
\caption{\label{P11}
The results for the
$\gamma^* p \rightarrow ~N(1440)P_{11}$
helicity amplitudes in units of $10^{-3}$GeV$^{-1/2}$.
For the DR and UIM results, the first and second uncertainties are, 
respectively,
the statistical uncertainty from the fit
and the model uncertainty, which
consists of uncertainties 
(I) (Sec. IV) and (II) (Sec. V) added in quadrature.
Final results are the average values of the amplitudes
found using DR and UIM; here the first 
uncertainty is statistical and the second one is the model
uncertainty discussed in Sec. VI. 
}
\end{table*}
 
\begin{table*}[t]
\begin{tabular}{|c|ccc|}
\hline
&&&\\
$Q^2$&DR&UIM&Final results\\
(GeV$^2$)&&&\\
&$A_{1/2}~~~~~~~~~~~~~~~~~~~S_{1/2}$
&$A_{1/2}~~~~~~~~~~~~~~~~~~~S_{1/2}$
&$A_{1/2}~~~~~~~~~~~~~~~~~~~S_{1/2}$\\
&&&\\
\hline
0.3&$89.4\pm 2.1\pm 1.3~~~~-11.0\pm 2.1\pm 0.9$
&$~~90.9\pm 2.3\pm 1.8~~~~-13.0\pm 2.2\pm 2.1$
&$~~~~~90.2\pm 2.2\pm 1.7~~~~-12.0\pm 2.2\pm 1.8$\\
0.4&$90.6\pm 1.7\pm 1.4~~~~-9.5\pm 1.9\pm 0.9$
&$~~92.9\pm 1.6\pm 2.2~~~~-15.9\pm 2.0\pm 2.2$
&$~~~~~91.8\pm 1.7\pm 2.1~~~~-12.7\pm 2.0\pm 3.6$\\
0.5&$90.5\pm 1.9\pm 1.6~~~~-10.8\pm 2.2\pm 1.2$
&$~~~91.7\pm 2.0\pm 2.7~~~~-16.7\pm 2.4\pm 2.4$
&$~~~~~91.1\pm 2.0\pm 2.2~~~~-13.8\pm 2.3\pm 3.5$\\
0.65&$90.0\pm 1.7\pm 1.8~~~~-12.9\pm 1.8\pm 1.0$
&$~~~~91.6\pm 1.8\pm 3.3~~~~-14.4\pm 1.9\pm 2.3$
&$~~~~~90.8\pm 1.8\pm 2.7~~~~-13.6\pm 1.9\pm 1.8$\\
0.9&$83.3\pm 2.4\pm 4.9~~~~-11.2\pm 3.8\pm 4.6$
&$~~~~~85.5\pm 2.3\pm 5.2~~~~-16.4\pm 3.9\pm 4.9$
&$~~~~~84.4\pm 2.4\pm 5.2~~~~-13.8\pm 3.9\pm 5.5$\\
1.72&$72.2\pm 1.5\pm 5.0~~~~-20.4\pm 1.8\pm 3.5$
&$~~~~~75.7\pm 1.4\pm 4.9~~~~-24.8\pm 1.6\pm 3.3$
&$~~~~~73.9\pm 1.5\pm 5.2~~~~-22.6\pm 1.7\pm 4.0$\\
2.05&$59.8\pm 1.6\pm 4.0~~~~-14.8\pm 2.0\pm 3.9$
&$~~~~~65.4\pm 1.7\pm 4.0~~~~-19.9\pm 1.9\pm 4.4$
&$~~~~~62.6\pm 1.7\pm 4.9~~~~-17.4\pm 1.9\pm 4.9$\\
2.44&$54.5\pm 2.1\pm 3.6~~~~-11.3\pm 2.7\pm 4.1$
&$~~~~~59.8\pm 2.2\pm 3.9~~~~-16.7\pm 2.9\pm 4.3$
&$~~~~~57.2\pm 2.2\pm 4.6~~~~-14.0\pm 2.8\pm 5.0$\\
2.91&$49.6\pm 2.0\pm 4.0~~~~~~-9.0\pm 2.6\pm 2.9$
&$~~~~~53.0\pm 1.9\pm 4.5~~~~-12.6\pm 2.8\pm 4.2$
&$~~~~~51.3\pm 2.0\pm 4.6~~~~-10.8\pm 2.7\pm 4.0$\\
3.48&$44.9\pm 2.2\pm 4.2~~~~~~-6.3\pm 3.2\pm 2.7$
&$~~~~~41.0\pm 2.4\pm 4.6~~~~-11.3\pm 3.4\pm 2.8$
&$~~~~~43.0\pm 2.3\pm 4.8~~~~-8.8\pm 3.3\pm 3.7$\\
4.16&$35.5\pm 3.8\pm 4.5~~~~~~-4.5\pm 6.2\pm 3.5$
&$~~~~~31.8\pm 3.6\pm 4.5~~~~-8.9\pm 5.9\pm 3.8$
&$~~~~~33.7\pm 3.7\pm 4.9~~~~-6.7\pm 6.0\pm 4.3$\\
\hline
\end{tabular}
\caption{\label{S11}
The results for the
$\gamma^* p \rightarrow ~N(1535)S_{11}$ 
helicity amplitudes in units of $10^{-3}$GeV$^{-1/2}$.
The amplitudes are extracted from the data on 
$\gamma^*p\rightarrow \pi N$  using
$\beta_{\pi N}(N(1535)S_{11})=0.485$ (see Subsection VII,C). 
The remaining legend is as for Table \ref{P11}.
}
\end{table*}

\begin{table*}[t]
\begin{tabular}{|c|cc|}
\hline
&&\\
$Q^2$&DR&UIM\\
(GeV$^2$)&&\\
&$A_{1/2}~~~~~~~~~~~~~~~~~~A_{3/2}~~~~~~~~~~~~~~~~~~~S_{1/2}$&
$A_{1/2}~~~~~~~~~~~~~~~~~~A_{3/2}~~~~~~~~~~~~~~~~~~~S_{1/2}$\\
&&\\
\hline
0.3&$-51.8\pm 1.9\pm 0.8~~77.2\pm 2.2\pm 0.7~-43.7\pm 2.4\pm 1.0 $
&$~~~~-54.1\pm 1.8\pm 1.8~~75.1\pm 2.2\pm 2.1~-48.4\pm 2.4\pm 2.3$ \\
0.4&$-57.0\pm 1.4\pm 0.9~~70.5\pm 1.8\pm 0.7~-39.7\pm 1.9\pm 1.0 $
&$~~~~-59.7\pm 2.1\pm 2.4~~67.6\pm 1.9\pm 2.2~-43.6\pm 2.1\pm 2.4$ \\
0.5&$-60.2\pm 2.0\pm 0.9~~56.9\pm 1.7\pm 0.8~-35.5\pm 2.5\pm 0.8 $
&$~~~~-60.6\pm 2.2\pm 2.5~~60.0\pm 1.9\pm 2.4~-39.4\pm 2.4\pm 2.8$\\
0.65&$-66.0\pm 1.6\pm 1.1~~52.0\pm 1.4\pm 0.8~-32.7\pm 2.1\pm 0.7 $
&$~~~~-64.5\pm 1.8\pm 2.7~~54.2\pm 1.6\pm 2.8~-37.5\pm 1.9\pm 2.5$\\
0.9&$-58.9\pm 2.4\pm 2.7~~44.8\pm 2.6\pm 2.8~-29.0\pm 3.3\pm 2.5$
&$~~~~-64.9\pm 2.2\pm 2.9~~44.1\pm 2.6\pm 3.1~-34.3\pm 3.1\pm 3.0$\\
1.72&$-42.4\pm 1.2\pm 3.2~~18.7\pm 1.2\pm 3.2~-11.8\pm 1.1\pm 3.1$
&$~~~~-38.8\pm 1.3\pm 3.9~~21.4\pm 1.2\pm 3.5~~-9.1\pm 1.0\pm 1.8$\\
2.05&$-37.3\pm 1.4\pm 2.1~~15.6\pm 1.5\pm 2.3~-9.6\pm 1.6\pm 2.8$
&$~~~~-39.7\pm 1.5\pm 3.2~~18.3\pm 1.6\pm 2.6~~~-6.8\pm 1.5\pm 1.9$\\
2.44&$-36.4\pm 1.3\pm 2.4~~~11.2\pm 1.6\pm 2.1~~-5.5\pm 1.8\pm 1.6$
&$~~~~-36.3\pm 1.4\pm 2.6~~13.4\pm 1.7\pm 1.9~~~-3.6\pm 1.9\pm 1.6$\\
2.91&$-32.8\pm 1.8\pm 2.6~~~~5.8\pm 2.1\pm 2.9~~-3.3\pm 2.0\pm 1.5$
&$~~~~-31.0\pm 1.9\pm 2.2~~9.6\pm 2.0\pm 2.7~~~-2.3\pm 2.1\pm 1.6$\\
3.48&$-22.4\pm 2.1\pm 2.7~~~~5.5\pm 2.0\pm 5.5~~-5.3\pm 2.5\pm 2.0$
&$~~~~-24.9\pm 2.2\pm 2.9~~~8.2\pm 2.2\pm 5.2~~~-2.6\pm 2.6\pm 2.4$\\
4.16&$-19.1\pm 3.9\pm 3.0~~~~6.4\pm 3.0\pm 7.5~~-2.6\pm 4.8\pm 3.0$
&$~~~~-20.9\pm 4.2\pm 3.2~~~4.6\pm 3.2\pm 6.9~~~-0.7\pm 4.6\pm 3.2$\\
\hline
\end{tabular}
\caption{\label{D13}
The results for the
$\gamma^* p \rightarrow ~N(1520)D_{13}$ 
helicity amplitudes in units of $10^{-3}$GeV$^{-1/2}$.
The remaining legend is as for Table \ref{P11}.
}
\end{table*}

\begin{table}[t]
\begin{tabular}{|l|ccc|}
\hline
&&&\\
$~~Q^2$&$A_{1/2}$&$A_{3/2}$&$S_{1/2}$\\
(GeV$^2$)&&&\\
&&&\\
\hline
   0.3&$  -52.9\pm 1.8\pm 1.7$
&$  76.1\pm 2.2\pm 1.7$&$ -46.1\pm 2.4\pm 2.9$\\
   0.4&$  -58.3\pm 1.8\pm 2.1$
&$  69.1\pm 1.8\pm 2.1$&$ -41.7\pm 2.0\pm 2.6$\\
   0.5&$  -60.4\pm 2.1\pm 1.7$
&$  58.5\pm 1.8\pm 2.2$&$ -37.5\pm 2.5\pm 2.7$\\
   0.65&$ -65.2\pm 1.7\pm 2.0$
&$  53.1\pm 1.5\pm 2.1$&$ -35.1\pm 2.0\pm 2.9$\\
   0.9&$  -61.9\pm 2.3\pm 4.1$
&$  44.4\pm 2.6\pm 3.0$&$ -31.6\pm 3.2\pm 3.8$\\
   1.72&$ -40.6\pm 1.2\pm 4.0$
&$  20.0\pm 1.2\pm 3.6$&$ -10.5\pm 1.0\pm 2.8$\\
   2.05&$ -38.5\pm 1.5\pm 2.9$
&$  17.0\pm 1.5\pm 2.8$&$  -8.2\pm 1.5\pm 2.7$\\
   2.44&$ -36.3\pm 1.3\pm 2.5$
&$  12.3\pm 1.7\pm 2.3$&$  -4.6\pm 1.8\pm 1.9$\\
   2.91&$ -31.9\pm 1.8\pm 2.6$
&$   7.7\pm 2.0\pm 3.4$&$  -2.8\pm 2.0\pm 1.6$\\
   3.48&$ -23.6\pm 2.2\pm 3.1$
&$   6.8\pm 2.1\pm 5.5$&$  -4.0\pm 2.5\pm 2.6$\\
   4.16&$ -20.0\pm 4.1\pm 3.2$
&$   5.5\pm 3.1\pm 7.3$&$  -1.6\pm 4.7\pm 3.2$\\
\hline
\end{tabular}
\caption{\label{D13_av}
The average values of the 
$\gamma^* p \rightarrow ~N(1520)D_{13}$ 
helicity amplitudes 
found using DR and UIM (in units of $10^{-3}$GeV$^{-1/2}$).  
The first
uncertainty is statistical, and the second one is the model
uncertainty discussed in Sec. VI.
}
\end{table}

The comparison with the data 
for $\frac{d\sigma}{d\Omega}$ 
and $A_{LT'}$ is made in terms of the structure functions
$\sigma_T+\epsilon\sigma_L$, $\sigma_{TT}$, $\sigma_{LT}$, 
$\sigma_{LT'}$
and their Legendre moments. 
They are defined in the following way:
\begin{eqnarray}
&&\frac{d\sigma}{d\Omega}=\sigma_T+\epsilon\sigma_L
+\epsilon\sigma_{TT}\cos{2\phi}\\
&&+\sqrt{2\epsilon(1+\epsilon)}\sigma_{LT}\cos{\phi}
+h\sqrt{2\epsilon(1-\epsilon)}\sigma_{LT'}\sin{\phi}\nonumber,
\end{eqnarray}
where $\frac{d\sigma}{d\Omega}$
is the differential cross section 
of the reaction $\gamma^* N\rightarrow N\pi$ in its c.m. system,
assuming that the virtual photon flux factor is
\begin{equation}
\Gamma=\frac{\alpha}{2\pi^2 Q^2}\frac{(W^2-m^2)E_f}{2mE_i}
\frac{1}{1-\epsilon}, \nonumber
\end{equation}
$E_i,~E_f$ are the initial and final electron energies
in the laboratory frame, and 
$\epsilon$ is the polarization
factor of the virtual photon. $\theta$ and $\phi$
are the polar and azimuthal angles of the pion
in the c.m. system of the reaction $\gamma^* N\rightarrow N\pi$, and $h$
is the electron helicity.
The longitudinally polarized beam asymmetry is related
to the structure function $\sigma_{LT'}$ by:
\begin{equation}
A_{LT'}=\frac{\sqrt{2\epsilon(1-\epsilon)}\sigma_{LT'}\sin{\phi}}
{\frac{d\sigma}{d\Omega}(h=0)}.
\end{equation}

For the longitudinal target asymmetry $A_t$
and beam-target asymmetry $A_{et}$ we use the
relations  presented
in detail in Ref. \cite{Biselli}, where the experimental
results on these observables are reported.
These relations express $A_t$ and $A_{et}$ through
the response functions defined in Ref. \cite{Response}.
 
The Legendre moments of structure functions are defined
as the coefficients in the expansion of these
functions over Legendre polynomials $P_l(\cos{\theta})$:

\begin{eqnarray}
\sigma_T(W,\cos {\theta})+ 
\epsilon \sigma_L(W,\cos{\theta})&&=\\
\sum_{l=0}^{n} &&
D_l^{T+L}(W)P_l(\cos{\theta}),\nonumber\\
\sigma_{LT}(W,\cos{\theta})= \sin{\theta} 
\sum_{l=0}^{n-1}&&D_l^{LT}(W) 
P_l(\cos{\theta}),\\
\sigma_{LT'}(W,\cos{\theta}) = \sin{\theta} \sum_{l=0}^{n-1}
&&D_l^{LT'}(W) P_l(\cos{\theta}),\\
\sigma_{TT}(W,\cos{\theta})= \sin^2{\theta}\sum_{l=0}^{n-2} 
&&D_l^{TT}(W) 
P_l(\cos{\theta}).
\end{eqnarray}

The Legendre moments allow us
to present a comparison of the results with the data
over all energies and angles in compact form.

The  Legendre moment $D_0^{T+L}$ represents the
$\cos{\theta}$ independent part of
$\sigma_T+\epsilon \sigma_L$,
which is related to the $\gamma^* N\rightarrow \pi N$
total cross section:
\begin{eqnarray}
&&D_0^{T+ L}=\frac{1}{4\pi}(\sigma^T_{tot}+\epsilon 
\sigma^L_{tot})\equiv \frac{|\bf{q}|}{K}
(\tilde{\sigma}_{tot}^T
+\epsilon \tilde{\sigma}_{tot}^L),\\ 
&&\tilde{\sigma}_{tot}^T=\tilde{\sigma}_{1/2}+\tilde{\sigma}_{3/2},
\nonumber\\
&&\tilde{\sigma}_{1/2}=
\sum_{l=0}^{\infty}(l+1)(|A_{l+}|^2+|A_{(l+1)-}|^2),\nonumber\\
&&\tilde{\sigma}_{3/2}=
\sum_{l=1}^{\infty}\frac{l}{4}(l+1)(l+2)(|B_{l+}|^2+|B_{(l+1)-}|^2),
\nonumber\\
&&\tilde{\sigma}_{tot}^L=\frac{Q^2}{\bf{k}^2}
\sum_{l=0}^{\infty}(l+1)^3(|S_{l+}|^2+|S_{(l+1)-}|^2).\nonumber
\end{eqnarray}
Here $\bf{q}$ and $\bf{k}$ are, respectively, the pion and virtual photon three-momenta
in the c.m. system
of the reaction $\gamma^* N\rightarrow \pi N$, $K=(W^2-m^2)/2W$, and
\begin{eqnarray}
&&A_{l+}=\frac{1}{2}\left[(l+2){E}_{l+}
+l{M}_{l+}\right], \\
&&B_{l+}={E}_{l+}-{M}_{l+},\nonumber\\
&&A_{(l+1)-}=\frac{1}{2}\left[
(l+2){M}_{(l+1)-}-l{E}_{(l+1)-}\right],\nonumber \\
&&B_{(l+1)-}={E}_{(l+1)-}+{M}_{(l+1)-}.\nonumber
\end{eqnarray}

The resonance structures
related to the resonances
$\Delta(1232)P_{33}$ and $N(1520)D_{13}$, $N(1535)S_{11}$
are revealed in
$D_0^{T+L}$ as enhancements.
It can be seen that with increasing $Q^2$,
the resonant structure near $1.5~$GeV 
becomes increasingly dominant in comparison with
the $\Delta(1232)$.
At $Q^2\geq 1.72~$GeV$^2$, there is a shoulder between the $\Delta$
and $1.5~$GeV peaks, which
is related to the large contribution
of the broad Roper resonance.
As can be seen from Table \ref{P11}, the transverse helicity
amplitude $A_{1/2}$ for $\gamma^* p\rightarrow N(1440)P_{11}$,
which is large and negative at $Q^2=0$ \cite{PDG},
crosses zero between $Q^2=0.4$ and $0.65~$GeV$^2$
and becomes large and positive
at $Q^2=1.72~$GeV$^2$. With
increasing $Q^2$, this amplitude drops smoothly in magnitude.

There are dips in the 
Legendre moment $D_2^{T+ L}$ 
that are caused by the $\Delta(1232)P_{33}$
and  $N(1520)D_{13}$, $N(1535)S_{11}$
resonances. They are related
to the following contributions to $D_2^{T}$:
\begin{equation}
D_2^{T}= 
-\frac{|\bf{q}|}{K}\left[4Re(A_{0+}A^*_{2-})+|M_{1+}|^2\right].
\end{equation}
When $Q^2$ grows the dip related to the $\Delta(1232)P_{33}$
resonance becomes smaller compared to that
near $1.5~$GeV. 

At $Q^2>1.72~$GeV$^2$, the relative values of
the dip in $D_2^{T+ L}$ and the enhancement in $D_0^{T+ L}$
near $1.5~$GeV,  
and the 
shoulder between the $\Delta$
and $1.5~$GeV peaks in $D_0^{T+ L}$,
remain approximately the same 
with increasing $Q^2$. 
Our analysis shows that this
is a manifestation of the
slow falloff of the $A_{1/2}$ helicity amplitudes
of the transitions $\gamma^* p \rightarrow ~$
$N(1440)P_{11}$,
$N(1535)S_{11}$, $N(1520)D_{13}$ for these $Q^2$.

The  enhancement in $D_0^{T+ L}$ and the dip in 
$D_0^{TT}$ in the $\Delta$ peak are mainly related
to the  
$M_{1+}^{3/2}$ amplitude 
of the 
$\gamma^* p \rightarrow ~\Delta(1232)P_{33}$
transition: 
\begin{eqnarray}
&&D_0^{T+ L}\approx 
2\frac{|\bf{q}|}{K}|M_{1+}|^2,\\
&&D_0^{TT}\approx -\frac{3}{2} 
\frac{|\bf{q}|}{K}|M_{1+}|^2.
\end{eqnarray}

In Figs. \ref{at}-\ref{aet_04},
we show the results for the target and double
spin asymmetries for $\vec{e}\vec{p}\rightarrow ep\pi^0$ \cite{Biselli}.
The inclusion of these data into the analysis resulted
in a smaller magnitude of the $S_{1/2}$ amplitude
for the Roper resonance, and also in the larger $A_{1/2}$ and
smaller $|S_{1/2}|$ amplitudes for the $\gamma^* p \rightarrow 
~N(1535)S_{11}$ transition. These data had minor impact
on the $\gamma^* p \rightarrow ~\Delta(1232)P_{33}$ and
$N(1520)D_{13}$ amplitudes.

\section{Comparison with theoretical predictions}

In Figs. \ref{delta} and \ref{p11}-\ref{d13}, we present 
our final results from Tables \ref{m1+}-\ref{S11} and \ref{D13_av};
they are average 
values of the amplitudes
extracted using DR and UIM.

\subsection{$\Delta(1232)P_{33}$ resonance}

\begin{figure*}[htp]
\begin{center}
\includegraphics[width=8.5cm]{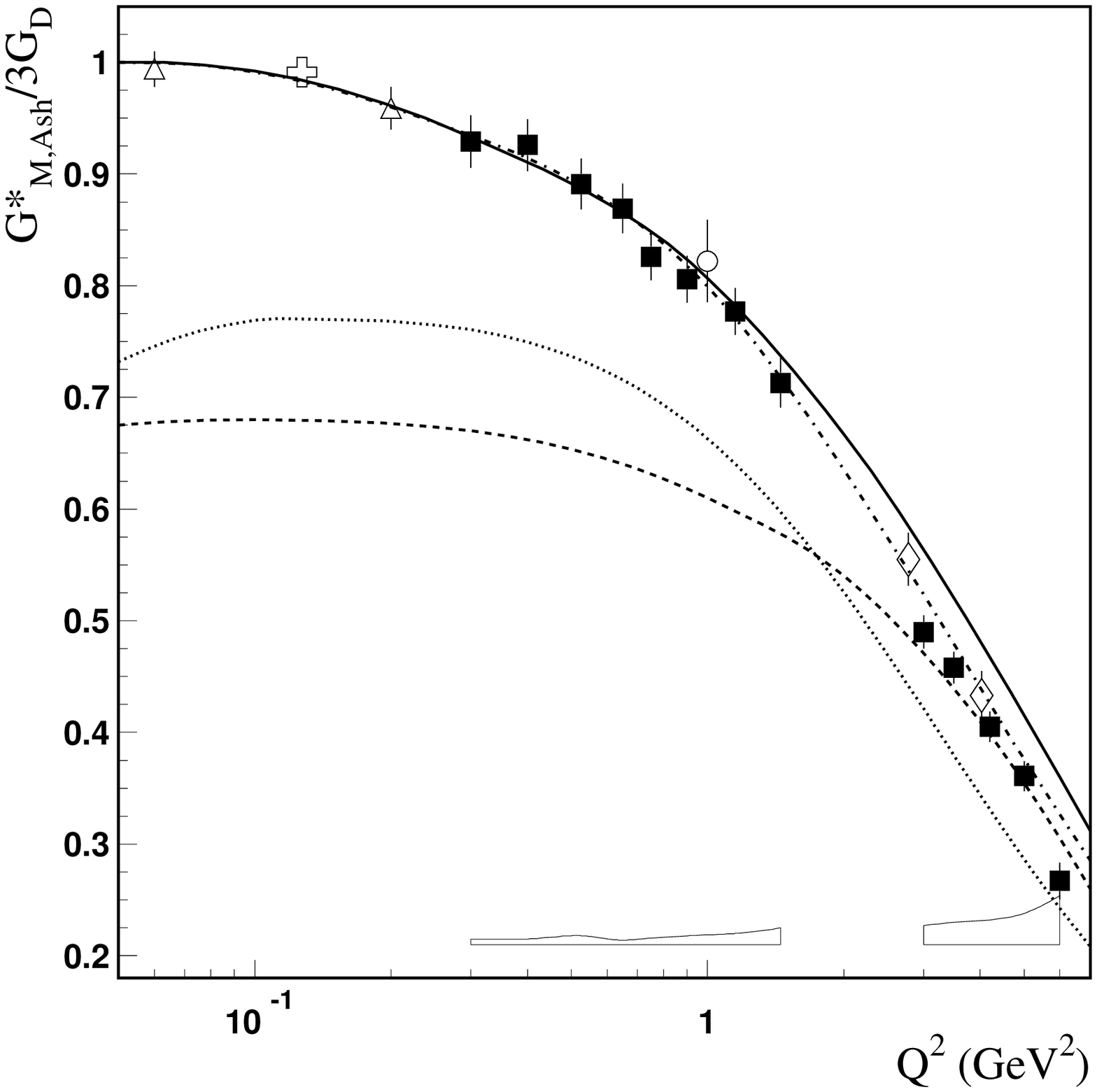}
\includegraphics[width=8.8cm]{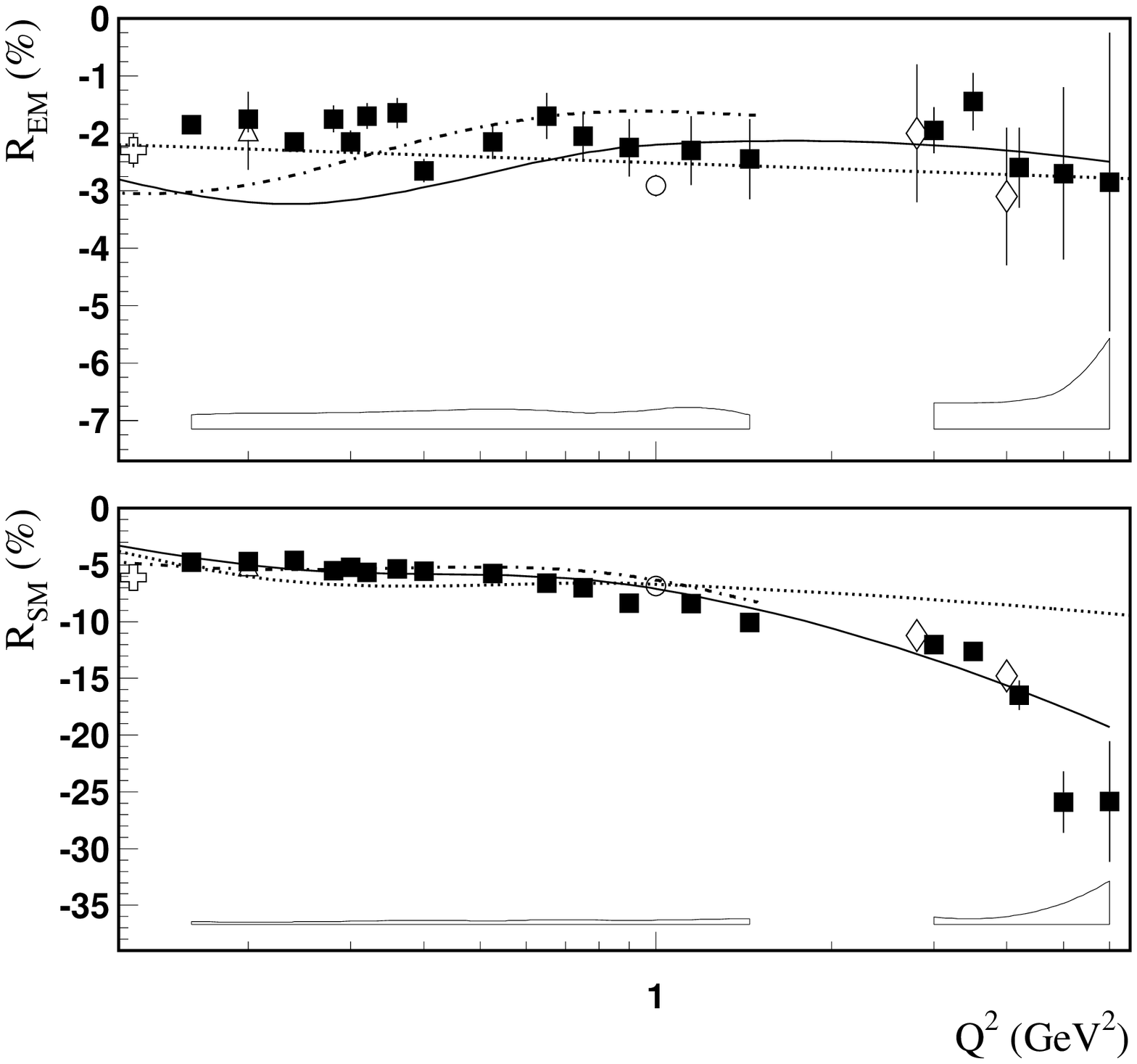}
\vspace{-0.1cm}
\caption{\small
Left panel: the form factor $G^*_{M}$
for the  $\gamma^* p \rightarrow~\Delta(1232)P_{33}$ 
transition relative to $3G_D$.
Right panel:
the ratios $R_{EM},~R_{SM}$.
The full boxes are
the results 
from Tables \ref{m1+}-\ref{RSM}
obtained in this work 
from CLAS data (Tables \ref{pol_data}, 
\ref{smith}, and \ref{pi0_data}).
The bands show the model uncertainties.
Also shown are the results from 
MAMI \cite{MAMI006,MAMI02} - open triangles, 
MIT/BATES \cite{BATES} - open crosses,
JLab/Hall C \cite{Frolov} - open rhombuses,
and JLab/Hall A \cite {KELLY1,KELLY2} - open circles.
The solid and dashed curves correspond to
the `dressed' and `bare' contributions from Ref. \cite{Sato};
for $R_{EM},~R_{SM}$, only the `dressed'
contributions are shown; the
`bare' contributions are close to zero.
The dashed-dotted curves are
the predictions obtained
in the large-$N_c$ limit of QCD \cite{GPD1,Pascalutsa}.
The dotted curve for $G^*_{M}$ is the prediction 
of a LF relativistic quark
model of Ref. \cite{Bruno}; 
the dotted curves for $R_{EM},~R_{SM}$
are the MAID2007 solutions \cite{MAID}.
\label{delta}}
\end{center}
\end{figure*}

The results for the
$\gamma^* p \rightarrow~\Delta(1232)P_{33}$ magnetic 
dipole form factor 
in the Ash convention \cite{Ash}
and for the ratios $R_{EM}\equiv E_{1+}^{3/2}/M_{1+}^{3/2}$,
$R_{SM}\equiv S_{1+}^{3/2}/M_{1+}^{3/2}$
are presented in Fig. \ref{delta}.
The relationship between $G^*_{M,Ash}(Q^2)$ and the 
corresponding multipole
amplitude is given by: 
\begin{equation}
G^*_{M,Ash}(Q^2)=\frac{m}{k_r}\sqrt{\frac{8q_r 
\Gamma}{3\alpha}}M_{1+}^{3/2}(Q^2,W=M),
\end{equation}
where $M=1232~$MeV and $\Gamma=118~$MeV are the mean values
of the mass and width of the $\Delta(1232)P_{33}$ (Table 
\ref{parameters}), $q_r,k_r$ are the pion and virtual photon
three-momenta, respectively, in the c.m. system of the reaction
$\gamma^* p
\rightarrow p\pi^0$ at the
$\Delta(1232)P_{33}$ resonance position, and $m$ is the nucleon mass. 
This definition is related to the definition of $G^*_M$ in
the Jones-Scadron convention \cite{Scadron} by:
\begin{equation}
G^*_{M,J-S}(Q^2)=G^*_{M,Ash}(Q^2)\sqrt{
1+\frac{Q^2}{(M+m)^2}}.
\end{equation}
The low $Q^2$ data from 
MAMI \cite{MAMI006,MAMI02} and 
MIT/BATES \cite{BATES}, and earlier
JLab Hall C \cite{Frolov} and
Hall A \cite {KELLY1,KELLY2} results
are also shown.
The form factor 
$G^*_{M}(Q^2)$ 
is presented relative to the dipole
form factor, which approximately describes the elastic
magnetic form factor of the proton. The plot shows that
new exclusive measurements of 
$G^*_{M}(Q^2)$, 
which now extend
over the range $Q^2=0.06-6~$GeV$^2$, confirm the rapid
falloff of
$G^*_{M}(Q^2)$ 
relative to the proton magnetic form factor
seen previously in inclusive measurements.

Fig. \ref{delta} shows
the long-standing discrepancy
between the measured  $G^*_{M}(Q^2)$ and 
the constituent quark model predictions;
here in comparison with the LF relativistic quark
model of Ref. \cite{Bruno}. 
Within  dynamical reaction models \cite{Yang,Kamalov,Sato,Lee},
the meson-cloud contribution
was identified as
the source of this discrepancy.
The importance of the pion (cloud) contribution
for the $\gamma^* p \rightarrow~\Delta(1232)P_{33}$
transition
is confirmed also by the lattice QCD calculations \cite{Alexandrou}.
In Fig. \ref{delta}, the results of the
dynamical model of Ref. \cite{Sato} are plotted. They show
the total amplitude (`dressed' form factor)
and the amplitude with the subtracted  meson-cloud
contribution (`bare' form factor). Very close results
are obtained within the dynamical model of Refs. \cite{Yang,Kamalov}.
The meson-cloud contribution
makes up more than 30\% of the total amplitude
at the photon point, and remains sizeable while $Q^2$ increases. 
                 
Figure \ref{delta} also shows the prediction
\cite{GPD1} obtained
in the large-$N_c$ limit of QCD, by relating
the $N\rightarrow \Delta$ and $N\rightarrow N$ GPDs.
A quantitative description of $G^*_M(Q^2)$
is obtained in the whole $Q^2$ range.

A consistent picture emerges from
the data for the ratios
$R_{EM}$ and $R_{SM}$: $R_{EM}$ remains negative, small and nearly
constant in the entire range $0<Q^2<6~$GeV$^2$; 
$R_{SM}$ remains negative, but its magnitude strongly
rises at high $Q^2$. It should be mentioned that the observed 
behavior of $R_{SM}$ at large $Q^2$ sharply disagrees
with the solution of MAID2007 \cite{MAID} based on
the same data set.
The magnitude of the relevant amplitude $S_{1+}^{3/2}$
can be directly checked using the data
for the structure function $\sigma_{LT}$, whose $\cos{\theta}$
behavior 
at $W=1.23~$GeV 
is dominated by the interference of this amplitude
with $M_{1+}^{3/2}$: 
\begin{equation}
D_1^{LT}(ep\rightarrow ep\pi^0)\approx\frac{8}{3}
\left(S_{1+}^{3/2}\right)^*M_{1+}^{3/2}.
\end{equation}

\begin{figure}[htp]
\begin{center}
\includegraphics[width=8.0cm]{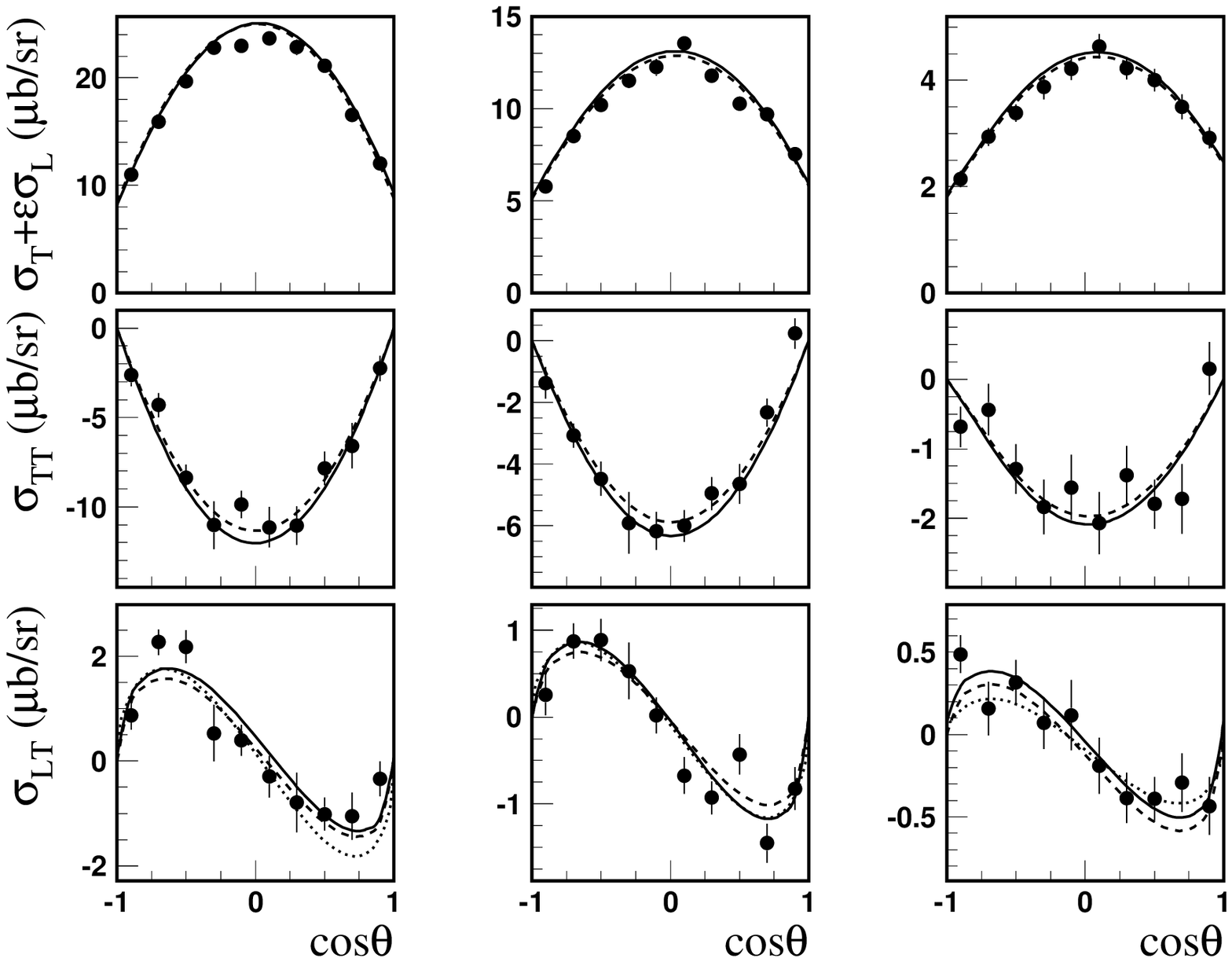}
\vspace{-0.1cm}
\caption{\small 
Our results for the
$ep\rightarrow ep\pi^0$ structure functions
(in $\mu$b/sr units)
in comparison with experimental data \cite{Joo1} for 
$W=1.23~$GeV.
The columns correspond to 
$Q^2=0.4,~0.75,~1.45~$GeV$^2$.
The solid (dashed) curves
correspond to the results obtained using DR (UIM) approach.
The dotted curves are from MAID2007 \cite{MAID}.
\label{str_04}}
\end{center}
\end{figure}

\begin{figure}[htp]
\begin{center}
\includegraphics[width=8.0cm]{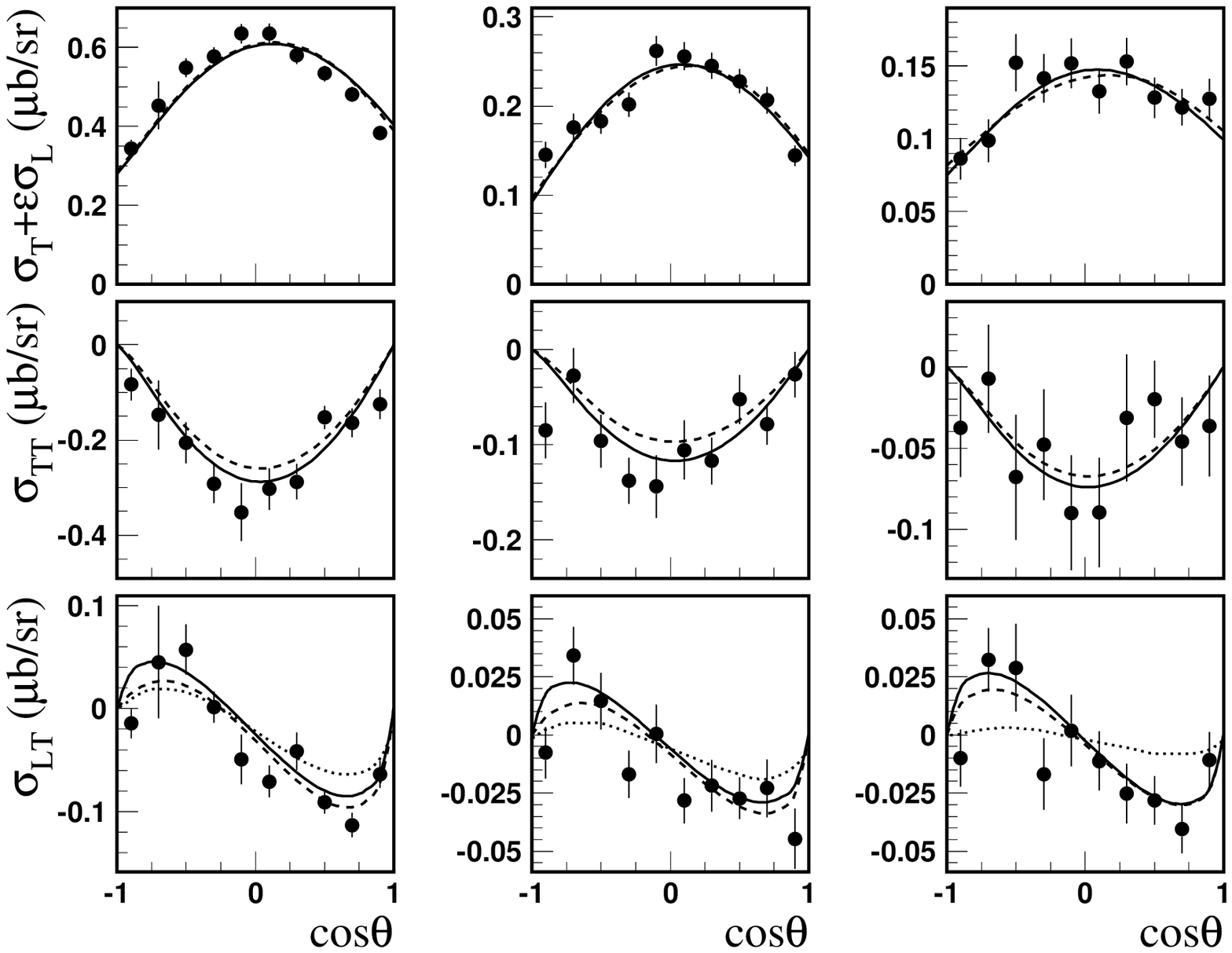}
\vspace{-0.1cm}
\caption{\small 
Our results for the
$ep\rightarrow ep\pi^0$ structure functions
(in $\mu$b/sr units)
in comparison with experimental data \cite{Ungaro} for 
$W=1.23~$GeV.
The columns correspond to 
$Q^2=3,~4.2,~5~$GeV$^2$.
The solid (dashed) curves
correspond to the results obtained using DR (UIM) approach.
The dotted curves are from MAID2007 \cite{MAID}.
\label{str_3}}
\end{center}
\end{figure}

The comparison of the experimental
data for the $ep\rightarrow ep\pi^0$ structure functions                                                  
with our results and the MAID2007 solution is shown
in Figs. \ref{str_04} and \ref{str_3}. 
At $Q^2=0.4-1.45~$GeV$^2$ (Fig. \ref{str_04}),
MAID2007 describes the angular behavior
of $\sigma_{LT}$.
However, it increasingly
underestimates the strong $\cos{\theta}$ dependence of this 
structure function
with rising $Q^2$, which is the direct consequence of the small
values of $R_{SM}$ in the 
MAID2007 solution. At $Q^2\geq 3~$GeV$^2$ this is
demonstrated in Fig. \ref{str_3}. In terms of $\chi^2$ per data point
for $\sigma_{LT}$ at $W=1.23~$GeV, the situation is 
presented in Table \ref{chi2}.

\begin{table}[t]
\begin{center}
\begin{tabular}{|l|ccc|}
\hline
&&&\\
$~~Q^2$&&$\chi^2/$d.p.&\\
(GeV$^2$)&&&\\
&$~~$DR$~~$&$~~$UIM$~~$&$~~$MAID2007$~~$\\
&&&\\
\hline
$~~$0.4&2.0&2.3&2.6\\
$~~$0.75&1.3&1.8&1.3\\
$~~$1.45&0.9&1.1&1.0\\
$~~$3&1.6&1.9&4.8\\
$~~$4.2&1.5&1.8&2.9\\
$~~$5&1.0&1.3&2.6\\
\hline
\end{tabular}
\caption{\label{chi2} 
Our results obtained within DR and UIM,
and the results of the MAID2007 solution
\cite{MAID} for $\chi^2$ per data point
for $\sigma_{LT}$ at $W=1.23~$GeV 
for $ep\rightarrow ep\pi^0$ data \cite{Joo1,Ungaro}.}
\end{center}
\end{table}

In constituent quark models,
the nonzero magnitude of $E_{1+}^{3/2}$ can arise
only due to a deformation of the $SU(6)$ spherical symmetry
in the N and (or) $\Delta(1232)$ wave functions.
In this connection it is interesting that both dynamical
models \cite{Sato,Yang} give practically zero
`bare' values for $R_{EM}$ (as well as for $R_{SM}$).
The entire $E_{1+}^{3/2}$ amplitude in these models is due to
the quadrupole deformation that arises through the
interaction of the photon with the meson cloud.

The knowledge of the $Q^2$ behavior
of the ratios $R_{EM},R_{SM}$ is of great interest
as a measure of the $Q^2$ scale where 
the asymptotic domain of QCD may set in
for this resonance transition.
In the pQCD asymptotics $R_{EM}\rightarrow 100\%$ and
$R_{SM}\rightarrow const$. The measured values
of $R_{EM},R_{SM}$ show that in the range
$Q^2< 6~$GeV$^2$, there is no sign of an approach to the 
asymptotic pQCD regime in either of these ratios.

\subsection{$N(1440)P_{11}$ resonance}

\begin{figure*}[htp]
\begin{center}
\includegraphics[width=11.0cm]{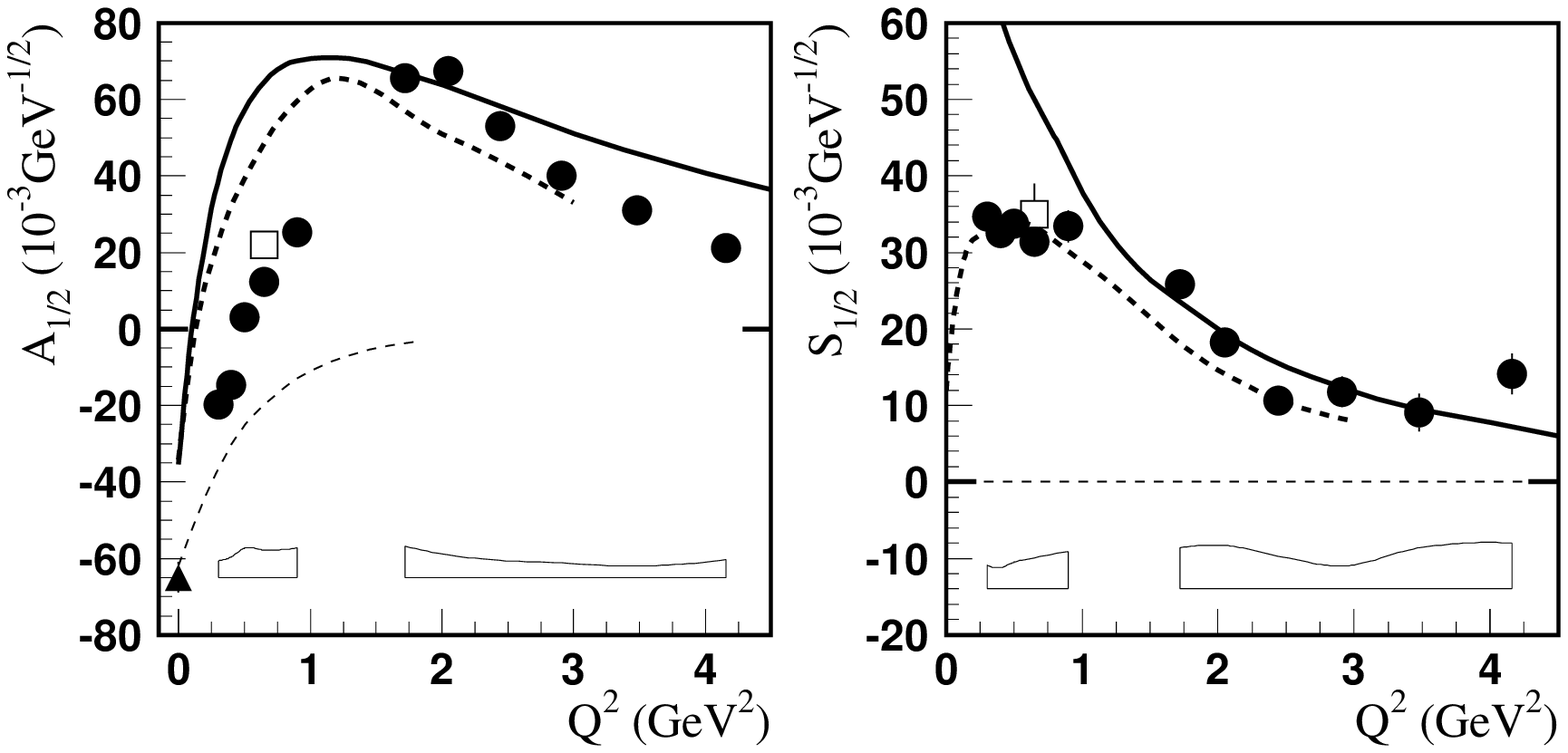}
\vspace{-0.1cm}
\caption{\small
Helicity amplitudes
for the  $\gamma^* p \rightarrow~N(1440)P_{11}$ 
transition.
The full circles are
the results from Table \ref{P11} obtained  
in this work from
CLAS data (Tables \ref{pol_data}-\ref{pi0_data}).
The bands show the model uncertainties.
The open boxes are  the results of the combined analysis
of CLAS single $\pi$ and 2$\pi$ electroproduction data 
\cite{Azn065}.
The full triangle at $Q^2=0$ is the RPP estimate \cite{PDG}.
The thick curves correspond to the results obtained in the
LF relativistic quark models 
assuming that $N(1440)P_{11}$
is a first radial excitation of the $3q$ ground state:  
\cite{Capstick} (dashed),
\cite{AznRoper} (solid).
The thin dashed curves are
obtained assuming that $N(1440)P_{11}$
is a gluonic baryon excitation (q$^3$G hybrid state) \cite{Li2}.
\label{p11}}
\end{center}
\end{figure*}

The results for the
$\gamma^* p \rightarrow~N(1440)P_{11}$ helicity 
amplitudes are presented in Fig. \ref{p11}.
The high $Q^2$ amplitudes ($Q^2=1.72-4.16~$GeV$^2$) and the
results for $Q^2=0.4,0.65~$GeV$^2$ were already
presented and discussed in Refs. \cite{Azn04,Roper}.
In the present paper the
data for $Q^2=0.4,0.65~$GeV$^2$ were reanalysed taking
into account the recent CLAS polarization measurements on the target and 
beam-target asymmetries \cite{Biselli}.
Also included are  new results
extracted at $Q^2=0.3,0.525,0.9~$GeV$^2$.    

By quantum numbers, the most natural classification of the
Roper resonance in the constituent quark model is a first
radial excitation of the $3q$ ground state. However, the difficulties
of quark models to describe the low mass and large
width of the $N(1440)P_{11}$,
and also its photocouplings
to the proton and neutron, gave rise to numerous
speculations. Alternative
descriptions of this state as a
gluonic baryon excitation \cite{Li1,Li2}, or a hadronic
N$\sigma$ molecule \cite{Krehl}, were suggested.
The CLAS measurements, for the first time, made possible
the determination of the electroexcitation
amplitudes of the Roper resonance on the proton
up to $Q^2=4.5~$GeV$^2$.
These results are crucial for
the understanding of the nature of this state.
There are several specific features in the extracted
$\gamma^* p \rightarrow~N(1440)P_{11}$ amplitudes
that are very important to test models.
First, the specific behavior of the transverse amplitude 
$A_{1/2}$,
which being large and negative at $Q^2=0$,
becomes large and positive at $Q^2\simeq 2~$GeV$^2$, and
then drops slowly with $Q^2$. Second,
the relative sign between the longitudinal $S_{1/2}$ and
transverse $A_{1/2}$ amplitudes.
And third,
the common sign of the amplitudes $A_{1/2},S_{1/2}$ extracted from the 
data
on $\gamma^*p\rightarrow \pi N$ includes signs
from the $\gamma^*p\rightarrow N(1440)P_{11}$
and $N(1440)P_{11}\rightarrow \pi N$ vertices;
both signs should be taken
into account while comparing with model predictions.
All these characteristics
are described by the
light-front relativistic quark models of Refs.
\cite{Capstick,AznRoper} assuming that
$N(1440)P_{11}$ is the first radial excitation of 
the $3q$ ground state.
Although the models \cite{Capstick,AznRoper}  fail to describe 
numerically the
data at small $Q^2$, this can have the natural explanation in the
meson-cloud contributions, which 
are expected to be large
for low $Q^2$ \cite{Lee1}.

\subsection{$N(1535)S_{11}$ resonance}

\begin{figure*}[htp]
\begin{center}
\includegraphics[width=11.0cm]{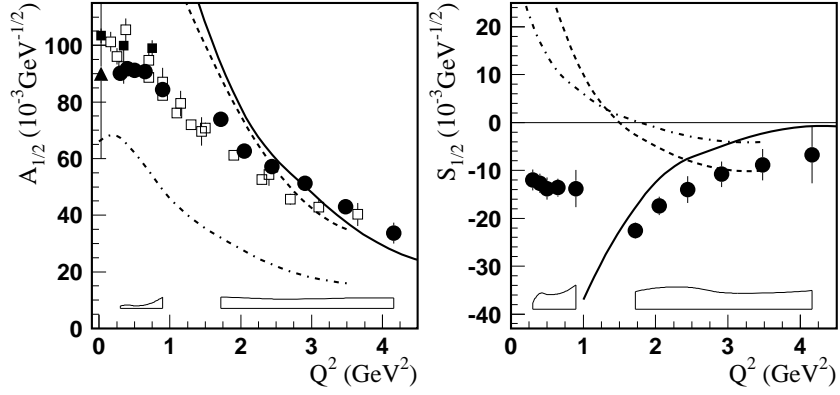}
\vspace{-0.1cm}
\caption{\small
Helicity amplitudes
for the  $\gamma^* p \rightarrow~N(1535)S_{11}$ 
transition. The legend is partly as for Fig. \ref{p11}.
The solid boxes are  the results extracted
from $\eta$ photo- and electroproduction data in Ref. \cite{Azneta},
the open boxes show the results from $\eta$  
electroproduction 
data \cite{Armstrong,Thompson,Denizli}.
The data are presented assuming 
$\beta_{\pi N}=0.485$,
$\beta_{\eta N}=0.46$
(see Subsection VII,C).
The results of the
LF relativistic quark models are given by the dashed  \cite{Capstick} 
and dashed-dotted \cite{Simula1} curves. The solid curves
are the central values of the amplitudes found within
light-cone sum rules using lattice results for light-cone
distribution amplitudes of the $N(1535)S_{11}$ resonance \cite{Braun}.
\label{s11}}
\end{center}
\end{figure*}

For the first time, the $\gamma^*N\rightarrow
N(1535)S_{11}$ transverse helicity amplitude
has been extracted from the $\pi$
electroproduction data
in a wide range of $Q^2$ (Fig.
\ref{s11}),
and the results confirm the 
$Q^2$-dependence of this amplitude
observed in $\eta$ electroproduction.
Numerical comparison of the  results
extracted from 
the  $\pi$ and $\eta$ photo- and electroproduction data
depends on the relation between
the branching ratios
to the $\pi N$ and $\eta N$ channels.
Consequently, it contains an arbitrariness connected
with the uncertainties of these branching ratios:
$\beta_{\pi N}=0.35-0.55$, $\beta_{\eta N}=0.45-0.6$ \cite{PDG}.
The amplitudes extracted from $\eta$ photo- and electroproduction
in Refs. \cite{Azneta,Armstrong,Thompson,Denizli}
correspond to $\beta_{\eta N}=0.55$.

The amplitudes found from
$\pi$ and $\eta$ data can be used to specify
the relation between
$\beta_{\pi N}$ and $\beta_{\eta N}$.
From the fit to these amplitudes
at $0\leq Q^2<4.5~$GeV$^2$,  
we found
\begin{equation}
\frac{\beta_{\eta N}}{ \beta_{\pi N}}=0.95\pm 0.03.
\end{equation}
Further, taking into account the branching
ratio to the $\pi\pi N$ channel 
$\beta_{\pi\pi N}=0.01-0.1$ \cite{PDG},
which accounts
practically for all channels different
from $\pi N$ and $\eta N$,  
we find 
\begin{eqnarray}
&&\beta_{\pi N}=0.485\pm 0.008\pm 0.023, \\
&&\beta_{\eta N}=0.460\pm 0.008\pm 0.022.
\end{eqnarray}
The first error corresponds to the fit error in Eq. (30)
and the second error is related to the uncertainty of 
$\beta_{\pi\pi N}$.
The results shown in Fig. \ref{s11} correspond to
$\beta_{\pi N}=0.485,~\beta_{\eta N}=0.46$. 

The CLAS data on $\pi$
electroproduction allowed the extraction
of the 
longitudinal helicity amplitude
for the $\gamma^*N\rightarrow
N(1535)S_{11}$ transition with good precision. 
These results are crucial for testing theoretical
models. It turned out that
at $Q^2<2~$GeV$^2$, the sign of $S_{1/2}$ is not described by the quark
models. 
Here it should be mentioned that 
quark model predictions for the 
relative signs between the $S_{1/2}$ and $A_{1/2},A_{3/2}$ amplitudes,
are presented for the transitions 
$\gamma^*N\rightarrow
N(1535)S_{11}$ and $N(1520)D_{13}$ (Figs. \ref{s11} and \ref{d13})
according to the investigation made in Ref. \cite{Definitions}.
Combined with the difficulties of quark models to
describe
the substantial
coupling of $N(1535)S_{11}$ to the $\eta N$ channel \cite{PDG}
and to strange particles \cite{Liu,Xie},
the difficulty 
in the description of the sign of $S_{1/2}$ can be indicative of
a large meson-cloud contribution
and (or) of additional
$q\bar{q}$ components
in this state \cite{An}.
Alternative representations of the $N(1535)S_{11}$
as a meson-baryon molecule have been also discussed
\cite{Weise,Nieves,Oset1,Lutz}.

\subsection{$N(1520)D_{13}$ resonance}

\begin{figure*}[htp]
\begin{center}
\includegraphics[width=16.8cm]{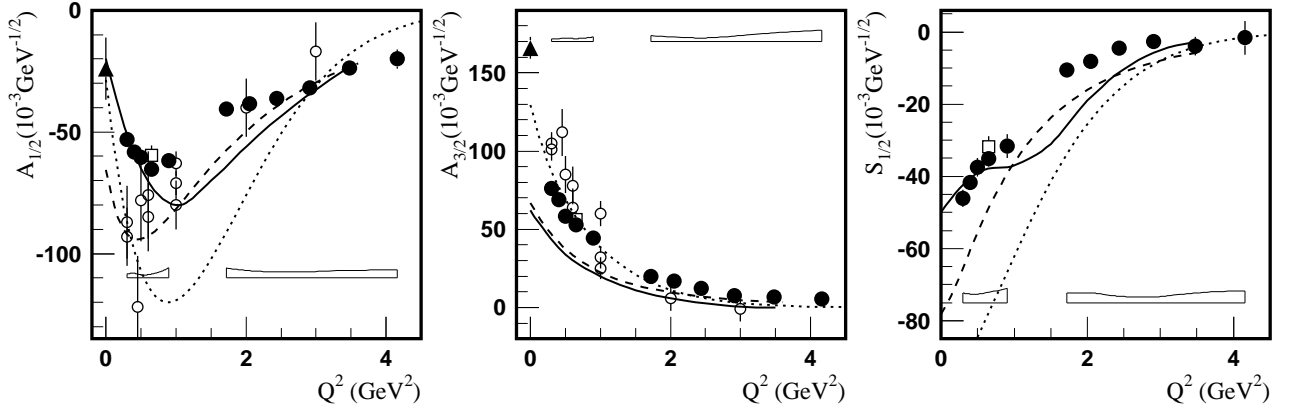}
\vspace{-0.1cm}
\caption{\small
Helicity amplitudes 
for the  $\gamma^* p \rightarrow~N(1520)D_{13}$ 
transition. The legend is partly as for Fig. \ref{p11}.
Open circles show the results
\cite{Foster} extracted from earlier  DESY 
\cite{Haidan,DESY}
and NINA \cite{NINA} data.
The curves correspond to the predictions of
the quark models:  
\cite{Warns} (solid), 
\cite{Santopinto} (dashed), 
and \cite{Merten} (dotted).
\label{d13}}
\end{center}
\end{figure*}

The results for the $\gamma^* p \rightarrow ~N(1520)D_{13}$ 
helicity amplitudes are shown in Fig. \ref{d13},
where the transverse amplitudes
are compared with those extracted
from earlier data.
The new data provide much more accurate results.
                                                                                
\begin{figure}[htp]
\begin{center}
\includegraphics[width=8.0cm]{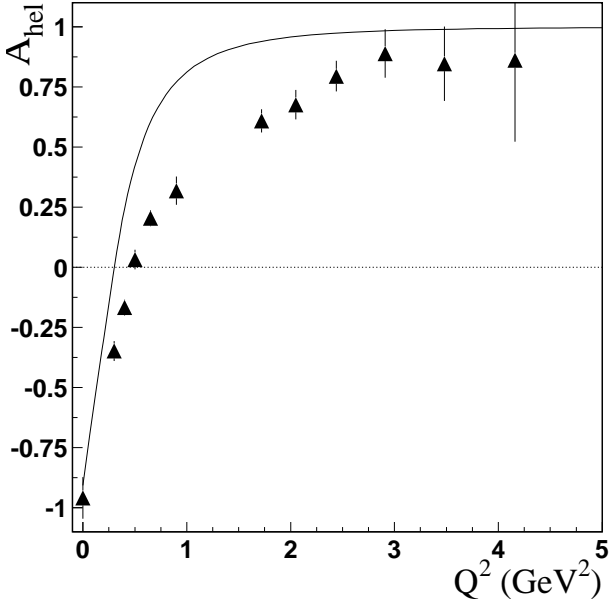}
\vspace{-0.1cm}
\caption{\small
The helicity asymmetry 
$A_{hel}\equiv (A^2_{1/2}-A^2_{3/2})/(A^2_{1/2}+A^2_{3/2})$ 
for the  $\gamma^* p \rightarrow~N(1520)D_{13}$ 
transition. 
Triangles show the results
obtained in this work.
The solid curve is the prediction of
the 
quark model with harmonic oscillator potential \cite{Isgur}.
\label{asym}}
\end{center}
\end{figure}
                                                                                
\begin{figure}[htp]
\begin{center}
\includegraphics[width=8.0cm]{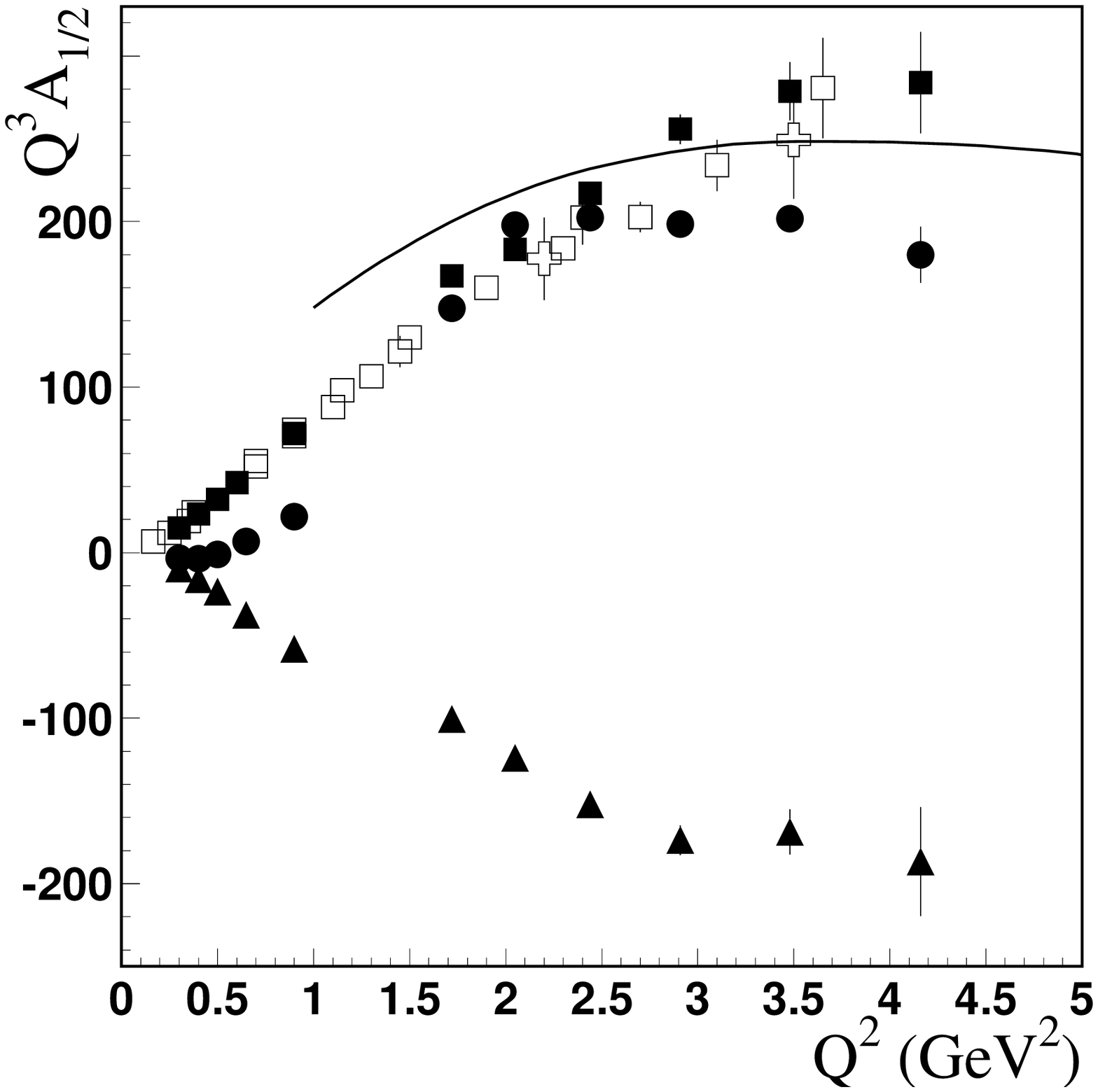}
\vspace{-0.1cm}
\caption{\small
The helicity amplitudes $A_{1/2}$ 
for the  $\gamma^* p \rightarrow~
N(1440)P_{11}$, 
$N(1520)D_{13}$, 
$N(1535)S_{11}$ 
transitions, multiplied by $Q^3$. 
The results
obtained in this work from the JLab-CLAS data on
pion electroproduction on the protons
are shown by solid circles ($N(1440)P_{11}$),
solid trangles ($N(1520)D_{13}$),
and solid boxes ($N(1535)S_{11}$).
Open boxes and crosses are the results for the $N(1535)S_{11}$ obtained
in $\eta$ electroproduction, respectively,
in HALL B \cite{Thompson,Denizli} and HALL C
\cite{Armstrong}.
The solid curve corresponds
to the amplitude  $A_{1/2}$ 
for the  $\gamma^* p \rightarrow~
N(1535)S_{11}$ transition 
found within
light-cone sum rules \cite{Braun}.
\label{asymptotics}}
\end{center}
\end{figure}

Sensitivity of the earlier data
to the 
$\gamma^* p \rightarrow~N(1520)D_{13}$
longitudinal helicity amplitude was limited.
The CLAS data allowed this 
amplitude to be determined with good precision and in a wide range of $Q^2$.

The obtained results show the rapid helicity switch from the dominance 
of the 
$A_{3/2}$ amplitude at the photon point to the dominance of $A_{1/2}$ at 
$Q^2>1~$GeV$^2$. 
This is demonstrated in Fig. \ref{asym} in terms of the helicity asymmetry.
Such behavior was predicted by a nonrelativistic
quark model with harmonic oscillator potential \cite{Close}.
Quark models also describe the sign and $Q^2$ dependence of the 
longitudinal amplitude. However, there are some shortcomings in the 
quark model description of the details of the $Q^2$ dependence of the 
$\gamma^* p \rightarrow~N(1520)D_{13}$ amplitudes.  
The amplitude $A_{3/2}$ is significantly underestimated
in all quark models 
for $Q^2<2~$GeV$^2$. Dynamical 
models predict large meson-cloud contributions to this
amplitude \cite{Lee1} 
that could explain the discrepancy.

Finally,  Fig. \ref{asymptotics}
shows the helicity amplitudes $A_{1/2}$ for the resonances
$N(1440)P_{11}$, $N(1520)D_{13}$,
$N(1535)S_{11}$, multiplied by $Q^3$. The data indicate that
starting with $Q^2=3~$GeV$^2$, these amplitudes
have a $Q^2$ dependence close to $1/Q^3$.
Such behaviour is expected  in pQCD in 
the limit $Q^2\rightarrow \infty$ \cite{Carlson}.
Measurements at higher
$Q^2$ are needed in order to check 
a possible $Q^3$ scaling of these amplitudes.

\section{Summary}

The electroexcitation amplitudes
for the low mass resonances
$\Delta(1232)P_{33}$, $N(1440)P_{11}$,
$N(1520)D_{13}$,
and $N(1535)S_{11}$ are determined in a wide range of $Q^2$
in the comprehensive analysis
of JLab-CLAS  data 
on differential cross sections, longitudinally polarized
beam asymmetries, and longitudinal target and beam-target asymmetries
for $\pi$ electroproduction off the proton.
A total of about 119,000 data points were included
covering the full azimuthal and polar angle range.
With this, we have complemented the previous analyses
\cite{Azn04,Azn065,Roper}
by including all JLab-CLAS pion electroproduction  data
available today.
We also have put significant effort into accounting for model
and systematic uncertainties of the extracted electroexcitation
amplitudes, by including the uncertainties
of hadronic parameters, such as
masses and widths of the resonances, 
the amplitudes of higher lying resonances,
the parameters which determine nonresonant contributions,
as well as the point-to-point systematics of the experimental
data and the overall normalization error of the
cross sections. 
Utilization of two approaches, DR and UIM,
allowed us to also estimate the model dependence of the results, which was
taken into account in the total model uncertainties of the extracted
amplitudes.

There are still additional uncertainties in the amplitudes presented in
this paper.
These are related to the lack of precise knowledge of the empirical
resonance
couplings to the $N\pi$ channel. However, we did not include these
uncertainties in the error budget as this is an overall multiplicative
correction that affects
all amplitudes for a given resonance equally, and, more importantly, the
amplitudes
can be corrected for these effects once improved hadronic couplings 
become
available.

The amplitudes for the electroexcitation of the $\Delta(1232)P_{33}$ 
resonance are determined in the range
$0.16 \leq Q^2\leq 6~$GeV$^2$.
The results are in agreement with the low $Q^2$ data from
MAMI \cite{MAMI006,MAMI02} and
MIT/BATES \cite{BATES}, and 
the JLab Hall A ($Q^2=1~$GeV$^2$) \cite {KELLY1,KELLY2} 
and  Hall C 
($Q^2= 2.8,4.2~$GeV$^2$) \cite{Frolov} data.

The results for the $\Delta(1232)P_{33}$ 
resonance show the importance of
the meson-cloud contribution to quantitatively explain the magnetic
dipole strength, as well as the electric and scalar quadrupole 
transitions.
They also do not show any tendency
of approaching the asymptotic QCD regime for
$Q^2\leq 6~$GeV$^2$. This was already mentioned in the
original paper \cite{Ungaro}, where the analysis was based
on the UIM approach only.
  
The amplitudes for the electroexcitation of the  
resonances $N(1440)P_{11}$,
$N(1520)D_{13}$, and $N(1535)S_{11}$ 
are determined in the range
$0.3 \leq Q^2< 4.5~$GeV$^2$.

For the Roper resonance, 
the high $Q^2$ amplitudes ($Q^2=1.7-4.5~$GeV$^2$) and the
results for $Q^2=0.4,0.65~$GeV$^2$ were already
presented and discussed in Refs. \cite{Azn04,Roper}.
In the present paper, the
data for $Q^2=0.4,0.65~$GeV$^2$ were reanalysed taking
into account the recent CLAS polarization measurements on the target and
beam-target asymmetries \cite{Biselli}.
Also included are  the new results
at $Q^2=0.3,0.525,0.9~$GeV$^2$.
The main conclusion for the Roper resonance is, as 
already reported in Ref. \cite{Roper},
that the data
on $\gamma^* p \rightarrow~N(1440)P_{11}$ 
available in the wide range of $Q^2$ 
provide a strong evidence for this
state to be predominantly the first radial excitation of the 3-quark
ground state. 

For the first time, the $\gamma^*p\rightarrow
N(1535)S_{11}$ transverse helicity amplitude
has been extracted from the $\pi$
electroproduction data
up to $Q^2=4.5~$GeV$^2$.
The results confirm the
$Q^2$-dependence of this amplitude
as observed in $\eta$ electroproduction.
The transverse amplitude found from
the $\pi$ and $\eta$ data allowed us to specify
the branching ratios
to the $\pi N$ and $\eta N$ channels for the $N(1535)S_{11}$.

Due to the CLAS measurements of $\pi$
electroproduction, for the first time
the $\gamma^*p\rightarrow
N(1520)D_{13}$ and $N(1535)S_{11}$ longitudinal helicity amplitudes
are determined from experimental data.
For the $\gamma^*p\rightarrow
N(1535)S_{11}$ transition,
the sign of $S_{1/2}$ is not 
described by quark models
at $Q^2<2~$GeV$^2$. 
Combined with the difficulties of quark models to
describe
the substantial
coupling of the $N(1535)S_{11}$ to the $\eta N$  
and strangeness channels,
this can be an indication of
a large meson-cloud contribution and/or of
additional $q\bar{q}$ components
in this state;
alternative representations of the $N(1535)S_{11}$
as a meson-baryon molecule are also possible.

The CLAS data provide much more accurate results
for the $\gamma^* p \rightarrow ~N(1520)D_{13}$
transverse helicity amplitudes 
than  those extracted
from earlier DESY and NINA data.
The data confirm the constituent quark model prediction
of the rapid helicity switch from the dominance of the
$A_{3/2}$ amplitude at the photon point to the dominance of $A_{1/2}$ at
$Q^2>1~$GeV$^2$.
Quark models also describe the sign and $Q^2$ dependence of the 
longitudinal amplitude. 

Starting with $Q^2=3~$GeV$^2$, the helicity amplitudes
$A_{1/2}$ for the resonances
$N(1440)P_{11}$,
$N(1520)D_{13}$, and $N(1535)S_{11}$ 
have a behaviour close to 
$1/Q^3$.
Measurements at higher
$Q^2$ are needed in order to check $Q^3$ scaling for these amplitudes.

\section{Acknowledgments}
This work was supported
in part by the U.S. Department of Energy and the National
Science Foundation, the Korea Research Foundation,
the French Commissariat a l'Energie Atomique
and CNRS/IN2P3, the
Italian Istituto Nazionale di Fisica Nucleare, 
the Skobeltsyn Institute of Nuclear Physics and
Physics Department at Moscow State University, and
the UK Science and Technology Facilities Research Council (STFC). 
Jefferson Science Associates, LLC, operates Jefferson Lab
under U.S. DOE contract DE-AC05-060R23177.

\end{document}